\documentclass[12pt,preprint]{aastex}
\usepackage[usenames]{color}

\shorttitle{Radio Drifting Pulsating Structures and Plasmoid Ejections}
\shortauthors{N. Nishizuka et al.}

\begin{document}

\title{PARTICLE ACCELERATION IN PLASMOID EJECTIONS DERIVED FROM RADIO DRIFTING PULSATING STRUCTURES}


\author{N. Nishizuka\altaffilmark{1}, M. Karlick\'{y}\altaffilmark{2}, and M. Janvier\altaffilmark{3}, M. B\'{a}rta\altaffilmark{2}}

\altaffiltext{1}{National Institute of Information and Communications Technology, 4-2-1, Nukui-Kitamachi, Koganei, Tokyo 184-8795, Japan; nishizuka.naoto@nict.go.jp}
\altaffiltext{2}{Astronomical Institute of the Academy of Sciences of the Czech Republic, 25165 Ond\v{r}ejov, Czech Republic}
\altaffiltext{3}{Department of Mathematics, University of Dundee, Dundee DD1 4HN, Scotland, United Kingdom}

\begin{abstract}
We report observations of slowly drifting pulsating structures (DPS) in the 0.8-4.5 GHz frequency range of the RT4 and RT5 radiospectrographs at Ond\v{r}ejov Observatory, 
between 2002 and 2012. We found 106 events of drifting pulsating structures, which we classified into 4 cases: (I) single events with a constant frequency drift [12 events], 
(II) multiple events occurring in the same flare with constant frequency drifts [11 events], (III) single or multiple events with increasing or decreasing frequency drift rates 
[52 events], and (IV) complex events containing multiple events occurring at the same time in the different frequency range [31 events]. Many DPSs are associated with hard 
X-ray bursts (15-25 keV) and soft X-ray gradient peaks, as they typically occurred at the beginning of the hard X-ray peaks. This indicates that DPS events are related to the 
processes of fast energy release and particle acceleration. Furthermore, interpreting DPSs as signatures of plasmoids, we measured their ejection velocity, their width and 
their height from the DPS spectra, from which we also estimated the reconnection rate and the plasma beta. In this interpretation, constant frequency drift indicates a 
constant velocity of a plasmoid, and an increasing/decreasing frequency drift indicates a deceleration/acceleration of a plasmoid ejection. The reconnection rate shows 
a good positive correlation with the plasmoid velocity. Finally we confirmed that some DPS events show plasmoid counterparts in AIA/SDO images.
\end{abstract}

\keywords{acceleration of particles --- magnetic reconnection --- Sun: flares --- Sun: particle emission --- Sun: radio radiation --- Sun: X-rays, gamma rays}

%
\section{Introduction}

Various plasma ejections in solar flares are observed in the solar corona. Ejections of soft X-ray emitting plasma blobs and hot loops, i.e. flux ropes in 3 dimensions, are called 
plasmoid ejections. Plasmoid ejections are observed in multi wavelength emissions: soft X-ray \citep[SXR: e.g.][]{shi95, ohy97, tsu97, nis10}, hard X-ray \citep[HXR: e.g.][]{hud01, 
sui03, gle13}, UV/EUV \citep[e.g.][]{liu10, tak12, kum13, liu13} and radio emissions \citep[e.g.][]{kli00, kun01, kha02, kar02, bai12}. Plasmoids gradually rise up in the preflare phase 
and are impulsively accelerated in the main phase, in association with HXR bursts. Similarly, when coronal mass ejections (CMEs) are accelerated in the main phase, both HXR bursts 
and a rapid increase of SXR emission are observed. \citep[CMEs:][]{zha01, tem08}. Some downward ejections of plasmoids, which collide with the loop-top region, are also reported 
\citep{kol07, liu10, mil10, tak12}. Plasmoid ejections are evidence of magnetic reconnection and multiple X-lines in a solar flare \citep[e.g.][]{shi95, ohy97, tsu97, kar04a, nis10}.

Accelerated electrons escaping from the reconnection regions in the upward direction are observed as type III bursts in radio emissions, while downward electron beams are 
observed as a reverse-drift burst \citep{asc97}. The observed frequency corresponds to the frequency of the plasma surrounding the electron beam, so that electrons propagating 
in a density gradient of the solar atmosphere show frequency shift in the radio spectra. Having these two bursts interpreted as bi-directional electron beams from one region, we 
can expect magnetic reconnection or acceleration in that location. Recently it has been shown that acceleration regions exist at several different heights during solar flares, 
both in observations \citep{asc02, kru10, rei11} and numerical simulations \citep{shim09, bar11, bar12, nisd13}.

Furthermore, some decimetric radio bursts in flares show coherent, quasi-periodic sequences of fast-drifting pulsed structures bound by a pair of slowly drifting low- and high- 
frequency cutoffs. These are called slowly drift pulsating structures \citep[DPSs: e.g.][]{kli00, kun01, kar02, kha02, kar04a, kar04b, tan07, tan08, tang08, bar08, wan12}. They are 
caused by quasi-periodic particle acceleration episodes that result from a highly dynamic regime of magnetic reconnection in an extended large-scale current sheet above the 
SXR loops, where reconnection is dominated by the repeated formation and the subsequent coalescence of plasmoids \citep{kli00, kar04a, bar08}. Electron beams trapped in a 
plasmoid are observed as pulsating structures in a DPS, while the motion of a plasmoid in a density gradient of the solar atmosphere is observed as a global frequency drift.

Electrons are accelerated in a DC field around the X-line in the vicinity of a plasmoid \citep{kar11}. The acceleration process of electrons forms an anisotropic velocity distribution, 
which excites the observed radio emission. The period of DPSs can be explained by the time scale of plasmoids determined by the tearing-mode oscillation \citep{tan07}. A few 
spectra of DPSs are shown to correspond to SXR emitting plasma ejecta \citep{kli00, kar04a, tan07}, in the flare ascending phase just after the onset. Previous studies have shown 
that all the DPS events are accompanied by GOES SXR flares \citep{tang08}, and they occur when the gradient of the GOES SXR curve reaches its maximum \citep{wan12}. In 
our 0.8-4.5 GHz frequency range, most of the DPSs take place at the beginning of the whole flare radio emission, i.e. before the type IV bursts, which is in this range a common 
flare burst type. From several very broad radio spectra taken by several radiospectrographs, Karlick\'{y} et al. (2002) found an association of DPSs with type II bursts, generated 
in the metric radio range by flare shocks. It agrees with the results of Tan et al. (2008) showing a close association of DPSs with CMEs or ejection events.

Plasmoid ejection is strongly related to the reconnection process, as shown in the ''plasmoid-induced reconnection model'' \citep{shi01}. When we assume an incompressible 
plasma around a current sheet, the conservation of mass fluxes of the inflow and the outflow leads to the following equation:
\begin{equation}
v_{in}=\frac{W_{pl}}{L_{in}}v_{pl}, 
\end{equation}
where $v_{in}$ is the inflow velocity, $L_{in}$ the length of the inflow region, $W_{pl}$ the width of the plasmoid, and $v_{pl}$ the plasmoid ejection velocity. This equation indicates 
that the plasmoid ejection induces a reconnection inflow, so that the ejection velocity has a good correlation with the inflow velocity. Even in the case of a compressible plasma, 
the inflow velocity and the net reconnection rate increase with the plasmoid ejection, when magnetic flux is piled up around a current sheet by the reconnection inflow at the 
same time as the plasmoid ejection \citep{ono11, hay12}.

Plasmoid ejections also have an important role in particle acceleration, as well as unsteady impulsive energy release. Current models of particle acceleration during solar flares 
considering plasmoid dynamics are as follows: (1) DC field acceleration in a reversed X-line between two colliding plasmoids \citep[e.g.][]{tan10, kar11}, (2) accelerations in a 
contracting plasmoid after coalescence \citep{dra06, tan10, oka10} and in multiple reconnection outflows by interacting plasmoids in 2 dimensions \citep{hos12}, and (3) 
acceleration at the fast shock above the loop-top where multiple plasmoids collide \citep{nis13}. Furthermore, turbulence produced by plasmoid dynamics may couple with 
stochastic accelerations.
   
In this paper, we classified DPSs into 4 cases with regards to the plasmoid dynamics, i.e. acceleration/deceleration, and estimated the reconnection rate by using DPSs data. The 
first advantage of our study is that we analyze 106 DPSs we found during the time period 2002-2012. The number of analyzed DPSs is larger than the previous studies and allows 
us to statistically study the sample. The second is that, in a gravitationally stratified atmosphere, DPSs show direct information along the vertical direction, i.e. the vertical ejection 
velocity and not the apparent velocity as usually deduced from imaging observations. The ejection direction is not always vertical, but in many cases the ejection may occur at some 
angle to the density gradient. This can affect the estimation of the ejection velocity from drift velocities in radio observations, though it would give an error of a factor $\sqrt{2}$ 
at most when the angle is 45$^{\circ}$. The third is that DPSs also contain information about electron beams, which allow us to compare the plasmoid ejection with the particle 
acceleration. This is why we also compared DPSs and HXR data, and discussed the relationship to the particle acceleration process. Finally we searched for plasmoid counterparts 
in imaging observations by the Solar Dynamics Observatory (SDO). We explain the analysis method in \S2, the analysis results in \S3, and we summarize our results and discuss 
them in \S4.

%
%
\section{Observations of DPS events and their Classification}

We used the radio spectrum data taken by the radiospectrograph at the Ond\v{r}ejov observatory \citep{jir93}. There are three radio telescopes (RT3, RT4 and RT5) in the frequency 
range of 0.8-4.5 GHz. The time cadence is 10 ms. The set of data we used covers 11 years from 2002 to 2012. The DPS events between 2002 and 2005 in our paper are partially 
the same as in a previous paper by B\'{a}rta et al. (2008), and we added some events during the same period with a more careful analysis and during 2006 and 2012 with new data 
(examples in Figures 1 and 2).

During this time period, there are lots of decimetric bursts observed, including type III or reverse-drift bursts and DPSs. The type III or reverse-drift bursts are detached broadband 
fast-drift bursts, whose typical durations are $<$1 s. Their origin is interpreted as free electron beams accelerated in the reconnection region. DPSs are ensembles of type III-like 
bursts, showing coherent, quasi-periodic sequence of fast drifting structures, with pulsed time structures of $<$1 s, usually bound by a constant or slowly drifting low- and high- 
frequency cutoffs. The fluctuations of DPSs in time may indicate intermittent energy release or acceleration. Fast drifting structures inside DPSs, interpreted as trapped particles 
inside plasmoids, are usually not U-shaped but straight. This is probably because electron beams inside plasmoids lose energy in very short time periods that are less than the circulation 
time due to the large density. The RT4 and RT5 observing range of 0.8-4.5 GHz corresponds to a plasmoid density of (0.04 - 3)$\times$10$^{11}$ cm$^{-3}$. Interpreting 
DPSs as signatures of plasmoids, the drifts towards lower/higher frequencies mean upward/downward motions of plasmoids. The DPSs with decreasing/increasing frequency drifts 
mean  acceleration/deceleration of plasmoids.
 
In this paper, we found 106 DPS events during 2002-2012. We dealt with multiple DPS events in a single flare as a single event, though B\'{a}rta et al. (2008) counted every DPSs 
as individual events. This is why the number of DPS events in B\'{a}rta et al. (2008) looks larger, although the number of DPSs we analyzed is much larger than in their paper. We 
summarized the events in Table 1, where we noted some characteristics of these DPSs. Some DPSs drift to lower frequencies, and others drift to higher frequencies or have no 
drift. The frequency drift sometimes varies in time. Some DPSs oscillate around a certain frequency. Most of the DPSs occurred at the beginning of the impulsive phase of flares 
and sometimes of the radio bursts known as broadband continuum (type IV burst), which is consistent with Tan, C. et al. (2008). Sometimes they appear after the impulsive or main 
phase, and small numbers of events are not associated with flares (inconsistent with Tang et al. 2007; this is discussed in detail in \S 3). 
   
\begin{figure}[hbtp]
\epsscale{.90}
\plotone{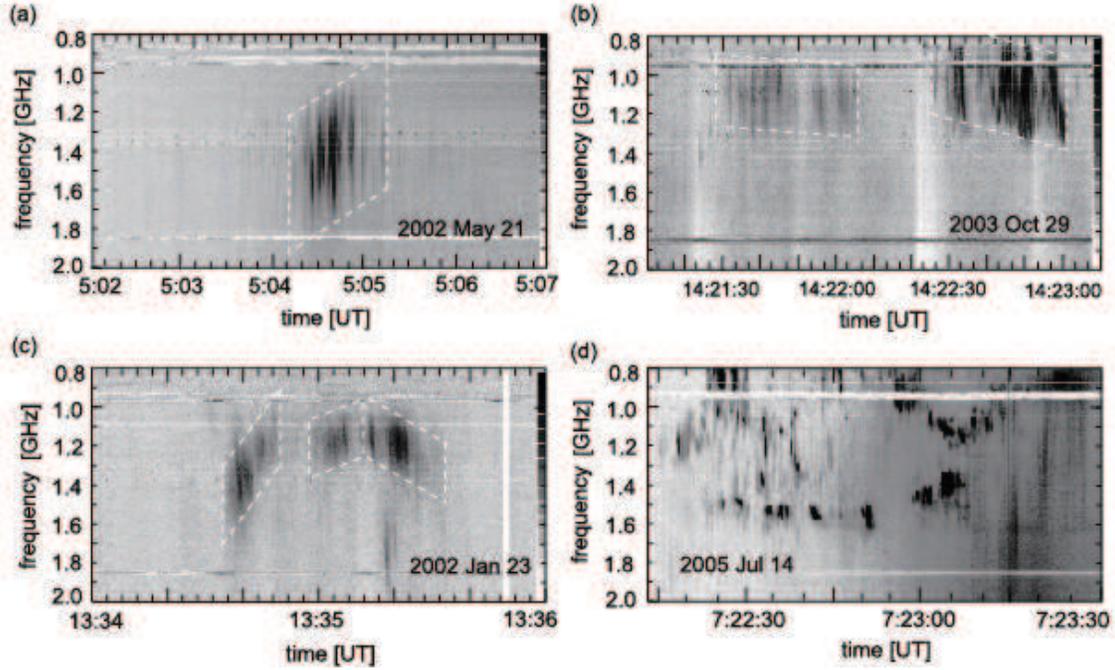}
\caption{Typical DPS events taken by Ond\v{r}ejov radiospectrograph telescopes: (a) (I) single events with constant frequency drift, (b) (II) multiple 
events with constant frequency drifts, (c) (III) single event or multiple events with increasing/decreasing frequency drift rates and (d) (IV) complex 
events including multiple events occurring at the same time in the different frequency range. \label{fig1}}
\end{figure}

\begin{figure}[hbtp]
\epsscale{.60}
\plotone{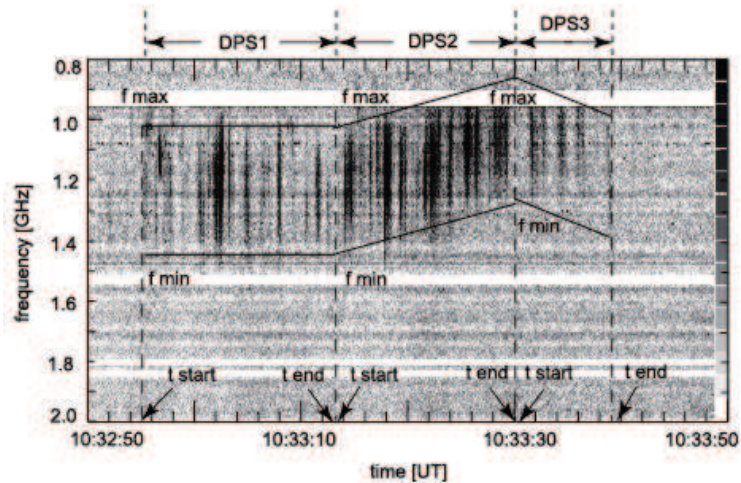}
\caption{A typical DPS event (case (III)) taken by Ond\v{r}ejov radiospectrograph telescope on 2011 Mar 6 with the measurable parameters indicated. \label{fig2}}
\end{figure}

First, we determine the background-subtracted radio spectra in the 0.8-4.5 GHz range and measure the start and end times, $t_{start}$ and $t_{end}$. We also measure 
the low- and high-frequency cutoffs of every decimetric bursts, listed as $f_{max}$ and $f_{min}$, at the start of the event as well as the end time (see Fig. 2). The upper/lower edge 
of the DPS is determined by eye. We determine the duration as $\Delta t$=$t_{end}-t_{start}$. It is noted that the frequency drift rate changes in quite a short time period $<$100 ms, 
as shown in Figure 2, so that the frequency drift rate shows discontinuity during the acceleration and in each period the frequency drift rate can be well fitted by a linear approximation 
(cases (III) and (IV) in the next paragraph). When a DPS event shows changes in the frequency drift rate, we separate the event to shorter periods with an approximately constant 
frequency drift and measure $t_{start}$, $t_{end}$, $f_{max}$ and $f_{min}$ for each period.

Following the different characteristics of DPSs, we classified our 106 DPS events into four categories: (I) single DPS events with a constant frequency drift (drift towards lower or 
higher frequencies), (II) multiple DPS events occurring in the same flare with constant frequency drifts, (III) single or multiple events with increasing or decreasing frequency drift 
rates and (IV) complex events containing multiple events occurring at the same time in the different frequency range. Cases (I) and (II) have only three events with no frequency 
drift, whereas cases (III) and (IV) include lots of events with no frequency drift preceding/following the acceleration/deceleration. The number of events for case (I) is 12 events, 
11 events for case (II), 52 events for case (III) and 31 events for case (IV). This leads to the fact that case (III) is the most common and case (IV) is the second. Therefore, DPSs 
usually have increasing or decreasing frequency drift rates. Those different categories are illustrated in Figure 1.

The motions of the DPSs in case (III) have typically three patterns: no drift to drift towards lower frequency, drift towards higher frequency to no drift, and drift towards higher 
frequency - no drift - drift towards lower frequency. It seems that most of the DPSs in case (III) tend to decrease the frequency drift rate. The initial frequency and the frequency 
drift rate of DPSs in case (II) are random. The patterns of the DPSs in case (IV) are sometimes very complex, composed by multiple DPSs with increasing/decreasing frequency drifts. 
Some of them are superposed and seem to be merged or split. Multiple DPSs appear at the same time in different frequency ranges. The observation of multiple DPSs 
may indicate multiple plasmoids formed in a current sheet via the secondary tearing mode instability, which leads to the repetition of the coalescence process and bursty ejections.

Figure 3 shows the year variation of the number of detected DPS events during 2002-2012. The first peak is around 2002, and the second one is in 2012. During 2006-2009, no 
DPSs were observed, except for one event in 2007. This seems to correspond with the 11-year period of solar activity variation. Especially case (III) shows the most sensitive and 
dramatic changes along the solar activity. 

\begin{figure}[hbtp]
\epsscale{.75}
\plotone{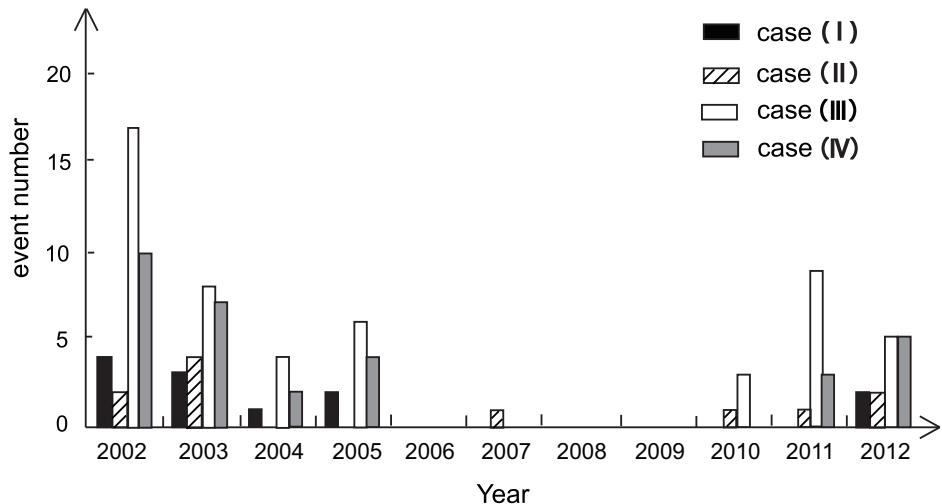}
\caption{Year variation of the number of DPS events during 2002-2012, with DPSs classified in four different categories. \label{fig3}}
\end{figure}

\clearpage

\begin{deluxetable}{llllllllllllrl}
\tabletypesize{\normalsize}
\tablecaption{Event list of the 106 DPS events during 2002-2012. \label{tbl1}}
\tablewidth{0pt}
\startdata
\tableline
\colhead{{\footnotesize Date}$^a$} & \colhead{{\footnotesize Time}$^b$} & \colhead{{\footnotesize Case}$^c$} & \colhead{{\tiny RHESSI}$^d$} & \colhead{{\tiny GOES}$^e$} & \colhead{{\footnotesize Date}$^a$} & \colhead{{\footnotesize Time}$^b$} & \colhead{{\footnotesize Case}$^c$} & \colhead{{\tiny RHESSI}$^d$} & \colhead{{\tiny GOES}$^e$}\\
\tableline
\colhead{{\footnotesize 2002/04/04}} & \colhead{{\footnotesize 15:27:40}} & \colhead{I} & \colhead{$\circ$} & \colhead{-} & \colhead{{\footnotesize 2002/05/17}} & \colhead{{\footnotesize 07:37:40}} & \colhead{I} & \colhead{$\circ$} & \colhead{-} & \colhead{}\\ 
\colhead{{\footnotesize 2002/05/21}} & \colhead{{\footnotesize 05:04:10}} & \colhead{I} & \colhead{-} & \colhead{$\circ$} & \colhead{{\footnotesize 2002/08/30}} & \colhead{{\footnotesize 13:27:37}} & \colhead{I} & \colhead{$\circ$} & \colhead{-} & \colhead{}\\ 
\colhead{{\footnotesize 2002/11/14}} & \colhead{{\footnotesize 11:09:26}} & \colhead{I} & \colhead{$\circ$} & \colhead{-} & \colhead{{\footnotesize 2003/01/26}} & \colhead{{\footnotesize 10:28:55}} & \colhead{I} & \colhead{-} & \colhead{$\circ$} & \colhead{}\\ 
\colhead{{\footnotesize 2003/05/27}} & \colhead{{\footnotesize 06:42:43}} & \colhead{I} & \colhead{-} & \colhead{$\circ$} & \colhead{{\footnotesize 2003/05/30}} & \colhead{{\footnotesize 08:51:55}} & \colhead{I} & \colhead{-} & \colhead{$\circ$} & \colhead{}\\
\colhead{{\footnotesize 2005/01/13}} & \colhead{{\footnotesize 11:25:32}} & \colhead{I} & \colhead{$\circ$} & \colhead{-} & \colhead{{\footnotesize 2005/06/14}} & \colhead{{\footnotesize 07:21:40}} & \colhead{I} & \colhead{-} & \colhead{$\circ$} & \colhead{}\\
\colhead{{\footnotesize 2012/07/06}} & \colhead{{\footnotesize 08:46:21}} & \colhead{I} & \colhead{-} & \colhead{$\times$} & \colhead{{\footnotesize 2012/07/11}} & \colhead{{\footnotesize 08:27:04}} & \colhead{I} & \colhead{-} & \colhead{$\circ$} & \colhead{}\\
\tableline
\colhead{{\footnotesize 2002/07/11}} & \colhead{{\footnotesize 14:46:00}} & \colhead{II} & \colhead{$\circ$} & \colhead{-} & \colhead{{\footnotesize 2002/11/10}} & \colhead{{\footnotesize 08:47:36}} & \colhead{II} & \colhead{$\circ$} & \colhead{-} & \colhead{}\\
\colhead{{\footnotesize 2003/06/02}} & \colhead{{\footnotesize 08:29:05}} & \colhead{II} & \colhead{-} & \colhead{$\circ$} & \colhead{{\footnotesize 2003/10/20}} & \colhead{{\footnotesize 07:09:54}} & \colhead{II} & \colhead{$\circ$} & \colhead{-} & \colhead{}\\
\colhead{{\footnotesize 2003/10/29}} & \colhead{{\footnotesize 14:21:30}} & \colhead{II} & \colhead{-} & \colhead{$\circ$} & \colhead{{\footnotesize 2003/11/11}} & \colhead{{\footnotesize 13:40:30}} & \colhead{II} & \colhead{-} & \colhead{$\circ$} & \colhead{}\\
\colhead{{\footnotesize 2007/01/10}} & \colhead{{\footnotesize 09:50:13}} & \colhead{II} & \colhead{-} & \colhead{$\circ$} & \colhead{{\footnotesize 2010/02/12}} & \colhead{{\footnotesize 11:24:09}} & \colhead{II} & \colhead{$\circ$} & \colhead{-} & \colhead{}\\
\colhead{{\footnotesize 2011/10/01}} & \colhead{{\footnotesize 12:37:35}} & \colhead{II} & \colhead{$\circ$} & \colhead{-} & \colhead{{\footnotesize 2012/04/25$^{\dagger}$}} & \colhead{{\footnotesize 08:56:00}} & \colhead{II} & \colhead{$\circ$} & \colhead{-} & \colhead{}\\
\colhead{{\footnotesize 2012/07/06}} & \colhead{{\footnotesize 06:14:02}} & \colhead{II} & \colhead{$\circ$} & \colhead{-} & \colhead{} & \colhead{} & \colhead{} & \colhead{} & \colhead{} & \colhead{}\\
\tableline
\colhead{{\footnotesize 2002/01/23}} & \colhead{{\footnotesize 13:34:36}} & \colhead{III} & \colhead{-} & \colhead{$\circ$} & \colhead{{\footnotesize 2002/02/07}} & \colhead{{\footnotesize 13:36:04}} & \colhead{III} & \colhead{-} & \colhead{$\circ$} & \colhead{}\\
\colhead{{\footnotesize 2002/04/10}} & \colhead{{\footnotesize 12:27:45}} & \colhead{III} & \colhead{$\circ$} & \colhead{-} & \colhead{{\footnotesize 2002/05/07$^{\ast}$}} & \colhead{{\footnotesize 14:08:33}} & \colhead{III} & \colhead{-} & \colhead{$\times$} & \colhead{}\\ 
\colhead{{\footnotesize 2002/05/20}} & \colhead{{\footnotesize 15:25:03}} & \colhead{III} & \colhead{-} & \colhead{$\circ$} & \colhead{{\footnotesize 2002/06/30}} & \colhead{{\footnotesize 11:01:13}} & \colhead{III} & \colhead{$\circ$} & \colhead{-} & \colhead{}\\ 
\colhead{{\footnotesize 2002/08/30$^{\ast}$}} & \colhead{{\footnotesize 13:38:45}} & \colhead{III} & \colhead{$\times$} & \colhead{-} & \colhead{{\footnotesize 2002/08/30}} & \colhead{{\footnotesize 14:36:35}} & \colhead{III} & \colhead{-} & \colhead{$\circ$} & \colhead{}\\
\colhead{{\footnotesize 2002/08/31}} & \colhead{{\footnotesize 10:01:27}} & \colhead{III} & \colhead{-} & \colhead{$\circ$} & \colhead{{\footnotesize 2002/08/31}} & \colhead{{\footnotesize 13:52:16}} & \colhead{III} & \colhead{-} & \colhead{$\circ$} & \colhead{}\\
\colhead{{\footnotesize 2002/08/31}} & \colhead{{\footnotesize 14:20:25}} & \colhead{III} & \colhead{$\circ$} & \colhead{-} & \colhead{{\footnotesize 2002/09/01}} & \colhead{{\footnotesize 12:13:55}} & \colhead{III} & \colhead{-} & \colhead{$\circ$} & \colhead{}\\
\colhead{{\footnotesize 2002/09/05}} & \colhead{{\footnotesize 16:24:55}} & \colhead{III} & \colhead{$\circ$} & \colhead{-} & \colhead{{\footnotesize 2002/09/20}} & \colhead{{\footnotesize 12:37:33}} & \colhead{III} & \colhead{$\circ$} & \colhead{-} & \colhead{}\\
\colhead{{\footnotesize 2002/09/21}} & \colhead{{\footnotesize 07:58:04}} & \colhead{III} & \colhead{$\circ$} & \colhead{-} & \colhead{{\footnotesize 2002/09/29}} & \colhead{{\footnotesize 06:38:40}} & \colhead{III} & \colhead{$\circ$} & \colhead{-} & \colhead{}\\
\colhead{{\footnotesize 2002/10/14}} & \colhead{{\footnotesize 09:29:22}} & \colhead{III} & \colhead{-} & \colhead{$\circ$} & \colhead{{\footnotesize 2003/03/18}} & \colhead{{\footnotesize 12:04:00}} & \colhead{III} & \colhead{$\circ$} & \colhead{-} & \colhead{}\\
\colhead{{\footnotesize 2003/04/30$^{\dagger}$}} & \colhead{{\footnotesize 07:30:14}} & \colhead{III} & \colhead{$\circ$} & \colhead{-} & \colhead{{\footnotesize 2003/05/27$^{\dagger}$}} & \colhead{{\footnotesize 08:01:20}} & \colhead{III} & \colhead{$\circ$} & \colhead{-} & \colhead{}\\
\colhead{{\footnotesize 2003/06/02$^{\dagger}$}} & \colhead{{\footnotesize 09:43:45}} & \colhead{III} & \colhead{$\circ$} & \colhead{-} & \colhead{{\footnotesize 2003/06/09}} & \colhead{{\footnotesize 10:46:16}} & \colhead{III} & \colhead{-} & \colhead{$\circ$} & \colhead{}\\
\colhead{{\footnotesize 2003/06/10}} & \colhead{{\footnotesize 08:35:35}} & \colhead{III} & \colhead{-} & \colhead{$\circ$} & \colhead{{\footnotesize 2003/10/25}} & \colhead{{\footnotesize 12:15:52}} & \colhead{III} & \colhead{-} & \colhead{$\circ$} & \colhead{}\\
\colhead{{\footnotesize 2003/11/18}} & \colhead{{\footnotesize 08:23:10}} & \colhead{III} & \colhead{$\circ$} & \colhead{-} & \colhead{{\footnotesize 2004/07/23}} & \colhead{{\footnotesize 06:43:42}} & \colhead{III} & \colhead{$\circ$} & \colhead{-} & \colhead{}\\
\colhead{{\footnotesize 2004/07/23}} & \colhead{{\footnotesize 17:19:06}} & \colhead{III} & \colhead{-} & \colhead{$\circ$} & \colhead{{\footnotesize 2004/07/25}} & \colhead{{\footnotesize 15:05:00}} & \colhead{III} & \colhead{$\circ$} & \colhead{-} & \colhead{}\\
\colhead{{\footnotesize 2004/10/30}} & \colhead{{\footnotesize 10:26:10}} & \colhead{III} & \colhead{$\circ$} & \colhead{-} & \colhead{{\footnotesize 2005/05/05}} & \colhead{{\footnotesize 14:25:55}} & \colhead{III} & \colhead{$\circ$} & \colhead{-} & \colhead{}\\ 
\colhead{{\footnotesize 2005/05/09}} & \colhead{{\footnotesize 09:55:40}} & \colhead{III} & \colhead{-} & \colhead{$\circ$} & \colhead{{\footnotesize 2005/05/10}} & \colhead{{\footnotesize 09:30:32}} & \colhead{III} & \colhead{$\circ$} & \colhead{-} & \colhead{}\\ 
\colhead{{\footnotesize 2005/05/11$^{\dagger}$}} & \colhead{{\footnotesize 15:35:07}} & \colhead{III} & \colhead{$\circ$} & \colhead{-} & \colhead{{\footnotesize 2005/07/12}} & \colhead{{\footnotesize 08:08:50}} & \colhead{III} & \colhead{$\circ$} & \colhead{-} & \colhead{}\\ 
\colhead{{\footnotesize 2005/09/13}} & \colhead{{\footnotesize 11:20:36}} & \colhead{III} & \colhead{$\circ$} & \colhead{-} & \colhead{{\footnotesize 2010/02/12}} & \colhead{{\footnotesize 11:22:00}} & \colhead{III} & \colhead{$\circ$} & \colhead{-} & \colhead{}\\ 
\colhead{{\footnotesize 2010/06/13}} & \colhead{{\footnotesize 05:35:00}} & \colhead{III} & \colhead{$\circ$} & \colhead{-} & \colhead{{\footnotesize 2010/07/30$^{\dagger}$}} & \colhead{{\footnotesize 11:08:50}} & \colhead{III} & \colhead{$\circ$} & \colhead{-} & \colhead{}\\ 
\colhead{{\footnotesize 2011/01/21}} & \colhead{{\footnotesize 10:50:32}} & \colhead{III} & \colhead{-} & \colhead{$\circ$} & \colhead{{\footnotesize 2011/02/19}} & \colhead{{\footnotesize 08:03:50}} & \colhead{III} & \colhead{$\circ$} & \colhead{-} & \colhead{}\\ 
\colhead{{\footnotesize 2011/03/06}} & \colhead{{\footnotesize 10:32:55}} & \colhead{III} & \colhead{-} & \colhead{$\circ$} & \colhead{{\footnotesize 2011/05/28}} & \colhead{{\footnotesize 11:16:11}} & \colhead{III} & \colhead{-} & \colhead{$\circ$} & \colhead{}\\ 
\colhead{{\footnotesize 2011/08/02}} & \colhead{{\footnotesize 11:37:56}} & \colhead{III} & \colhead{$\circ$} & \colhead{-} & \colhead{{\footnotesize 2011/08/08}} & \colhead{{\footnotesize 12:16:22}} & \colhead{III} & \colhead{$\circ$} & \colhead{-} & \colhead{}\\ 
\colhead{{\footnotesize 2011/08/08}} & \colhead{{\footnotesize 15:34:47}} & \colhead{III} & \colhead{$\circ$} & \colhead{-} & \colhead{{\footnotesize 2011/08/08}} & \colhead{{\footnotesize 15:56:24}} & \colhead{III} & \colhead{-} & \colhead{$\circ$} & \colhead{}\\ 
\colhead{{\footnotesize 2011/12/21}} & \colhead{{\footnotesize 11:03:44}} & \colhead{III} & \colhead{$\circ$} & \colhead{-} & \colhead{{\footnotesize 2012/05/08}} & \colhead{{\footnotesize 13:04:42}} & \colhead{III} & \colhead{-} & \colhead{$\circ$} & \colhead{}\\
\colhead{{\footnotesize 2012/05/09}} & \colhead{{\footnotesize 10:26:21}} & \colhead{III} & \colhead{-} & \colhead{$\circ$} & \colhead{{\footnotesize 2012/06/13}} & \colhead{{\footnotesize 12:50:04}} & \colhead{III} & \colhead{-} & \colhead{$\circ$} & \colhead{}\\
\colhead{{\footnotesize 2012/07/04}} & \colhead{{\footnotesize 14:39:46}} & \colhead{III} & \colhead{-} & \colhead{$\circ$} & \colhead{{\footnotesize 2012/07/12}} & \colhead{{\footnotesize 14:16:06}} & \colhead{III} & \colhead{-} & \colhead{$\circ$} & \colhead{}\\ 
\tableline
\colhead{{\footnotesize 2002/03/16}} & \colhead{{\footnotesize 11:22:20}} & \colhead{IV} & \colhead{-} & \colhead{$\circ$} & \colhead{{\footnotesize 2002/04/10}} & \colhead{{\footnotesize 06:56:26}} & \colhead{IV} & \colhead{-} & \colhead{$\times$} & \colhead{}\\
\colhead{{\footnotesize 2002/06/02}} & \colhead{{\footnotesize 10:44:40}} & \colhead{IV} & \colhead{$\times$} & \colhead{-} & \colhead{{\footnotesize 2002/08/17}} & \colhead{{\footnotesize 14:36:30}} & \colhead{IV} & \colhead{$\circ$} & \colhead{-} & \colhead{}\\
\colhead{{\footnotesize 2002/08/20}} & \colhead{{\footnotesize 14:28:30}} & \colhead{IV} & \colhead{-} & \colhead{$\circ$} & \colhead{{\footnotesize 2002/08/29}} & \colhead{{\footnotesize 12:46:50}} & \colhead{IV} & \colhead{-} & \colhead{$\circ$} & \colhead{}\\ 
\colhead{{\footnotesize 2002/09/17}} & \colhead{{\footnotesize 09:17:02}} & \colhead{IV} & \colhead{$\circ$} & \colhead{-} & \colhead{{\footnotesize 2002/09/28}} & \colhead{{\footnotesize 10:34:40}} & \colhead{IV} & \colhead{-} & \colhead{$\circ$} & \colhead{}\\ 
\colhead{{\footnotesize 2002/11/21}} & \colhead{{\footnotesize 12:43:18}} & \colhead{IV} & \colhead{$\circ$} & \colhead{-} & \colhead{{\footnotesize 2002/11/25}} & \colhead{{\footnotesize 10:41:49}} & \colhead{IV} & \colhead{$\circ$} & \colhead{-} & \colhead{}\\ 
\colhead{{\footnotesize 2003/01/03$^{\dagger}$}} & \colhead{{\footnotesize 13:25:27}} & \colhead{IV} & \colhead{$\circ$} & \colhead{-} & \colhead{{\footnotesize 2003/02/14}} & \colhead{{\footnotesize 09:16:06}} & \colhead{IV} & \colhead{$\circ$} & \colhead{-} & \colhead{}\\ 
\colhead{{\footnotesize 2003/05/28$^{\ast}$}} & \colhead{{\footnotesize 08:08:24}} & \colhead{IV} & \colhead{$\times$} & \colhead{-} & \colhead{{\footnotesize 2003/06/09}} & \colhead{{\footnotesize 16:27:01}} & \colhead{IV} & \colhead{$\circ$} & \colhead{-} & \colhead{}\\ 
\colhead{{\footnotesize 2003/06/10}} & \colhead{{\footnotesize 16:29:30}} & \colhead{IV} & \colhead{$\circ$} & \colhead{-} & \colhead{{\footnotesize 2003/06/11}} & \colhead{{\footnotesize 09:55:30}} & \colhead{IV} & \colhead{-} & \colhead{$\circ$} & \colhead{}\\
\colhead{{\footnotesize 2003/11/17}} & \colhead{{\footnotesize 09:20:10}} & \colhead{IV} & \colhead{-} & \colhead{$\times$} & \colhead{{\footnotesize 2004/01/08$^{\ast}$}} & \colhead{{\footnotesize 10:27:18}} & \colhead{IV} & \colhead{-} & \colhead{$\times$} & \colhead{}\\
\colhead{{\footnotesize 2004/10/30}} & \colhead{{\footnotesize 13:46:55}} & \colhead{IV} & \colhead{$\circ$} & \colhead{-} & \colhead{{\footnotesize 2005/05/09}} & \colhead{{\footnotesize 11:28:56}} & \colhead{IV} & \colhead{-} & \colhead{$\circ$} & \colhead{}\\
\colhead{{\footnotesize 2005/06/14}} & \colhead{{\footnotesize 07:52:27}} & \colhead{IV} & \colhead{$\times$} & \colhead{-} & \colhead{{\footnotesize 2005/07/11}} & \colhead{{\footnotesize 16:35:20}} & \colhead{IV} & \colhead{-} & \colhead{$\circ$} & \colhead{}\\
\colhead{{\footnotesize 2005/07/14}} & \colhead{{\footnotesize 07:22:19}} & \colhead{IV} & \colhead{-} & \colhead{$\circ$} & \colhead{{\footnotesize 2011/05/29}} & \colhead{{\footnotesize 09:45:45}} & \colhead{IV} & \colhead{-} & \colhead{$\circ$} & \colhead{}\\ 
\colhead{{\footnotesize 2011/09/24$^{\ast}$}} & \colhead{{\footnotesize 15:12:35}} & \colhead{IV} & \colhead{-} & \colhead{$\times$} & \colhead{{\footnotesize 2011/10/13}} & \colhead{{\footnotesize 14:50:22}} & \colhead{IV} & \colhead{-} & \colhead{$\circ$} & \colhead{}\\
\colhead{{\footnotesize 2012/01/14}} & \colhead{{\footnotesize 13:15:50}} & \colhead{IV} & \colhead{-} & \colhead{$\circ$} & \colhead{{\footnotesize 2012/07/03}} & \colhead{{\footnotesize 13:57:48}} & \colhead{IV} & \colhead{-} & \colhead{$\circ$} & \colhead{}\\
\colhead{{\footnotesize 2012/07/03}} & \colhead{{\footnotesize 17:00:18}} & \colhead{IV} & \colhead{$\times$} & \colhead{-} & \colhead{{\footnotesize 2012/07/06}} & \colhead{{\footnotesize 07:05:48}} & \colhead{IV} & \colhead{-} & \colhead{$\circ$} & \colhead{}\\ 
\colhead{{\footnotesize 2012/07/06}} & \colhead{{\footnotesize 13:29:25}} & \colhead{IV} & \colhead{-} & \colhead{$\circ$} &  \colhead{} & \colhead{} & \colhead{} & \colhead{} & \colhead{} & \colhead{}\\
\tableline
\enddata
\tablecomments{$^a$Date (yyyy/mm/dd), $^b$Start time [UT], $^c$Classification: case (I)-(IV), $^d$RHESSI peak (with/without $\circ$/$\times$; no data $-$), $^e$GOES time derivative peak 
(with/without $\circ$/$\times$, events with RHESSI data are skipped $-$). Here we consider events occurring during different flares. $\dagger$ mark denotes the events with 
HXR peaks but without GOES SXR flares. $\ast$ mark denotes the events with neither HXR nor GOES SXR flares.}
\end{deluxetable}

\clearpage
%
%
\section{DPSs in association with HXR bursts}
\subsection{Measurements of HXR peaks} 

DPSs  are ensembles of type III-like bursts, i.e. emissions from high energy electron beams via wave-particle interactions. In DPSs, electrons are trapped in plasmoids. Therefore these 
trapped electrons cannot reach dense layers of the solar atmosphere where the X-rays are usually generated, i.e. the chromosphere and the photosphere. However, there are electrons 
accelerated simultaneously in the same reconnection region, which are not trapped in the plasmoids. These can then generate HXRs in deep and dense layers of the 
Sun's atmosphere. Therefore, some association of DPSs and HXR bursts can be expected \citep{kar04b}. This association can be further 
influenced by the fact that any beam of superthermal electrons generate the X-rays in dense layers, but the radio emission of the beam can be reduced due to an inappropriate beam 
distribution function, reduced wave conversion, and absorption effects. On the other hand, if the magnetic field structure associated with many interacting plasmoids becomes complicated, 
all accelerated electrons are trapped in this structure and do not reach the dense layers of the solar atmosphere. In this case we can observe DPSs without corresponding HXRs. Since 
a DPS (a series of acceleration episode) lasts much longer than a single type III bursts (one acceleration episode), we study here the mean association of DPSs with HXR peaks, i. e. 
not for individual acceleration episodes as in the analysis of type III bursts and HXR peaks \citep{ben05}. 

To compare DPS events with HXRs, we used data taken by RHESSI \citep{lin02}, whose time cadence is 4 s. We checked all the RHESSI data for the 106 DPSs presented above, and 51\% 
of all events (54 DPSs) turned out to be simultaneously observed with RHESSI. For these events, we drew light curves integrated on-disk in three different energy ranges, 15-25 keV, 
25-50 keV and 50-100 keV. For one DPS event not correlated with any RHESSI data, we used data taken by the Czech-made Hard X-Ray Spectrometer \citep[HXRS][]{far01} onboard 
the U.S. Department of Energy Multispectral Thermal Imager satellite (MTI). HXRS has three energy bands: HXRS-S1 for 19-29 keV, HXRS-S2 for 44-67 keV and HXRS-S3 for 100-147 
keV, with a time cadence of 200 ms. For the other events with neither RHESSI nor HXRS data, we calculated the time derivative of SXR light curves taken by GOES, whose peak is believed 
to be well correlated with HXR peaks, as known from the Neupert effect (1968). The criterion used to distinguish whether there is a HXR component is an evident increase 
of HXR emission up to over 40 counts with 4 s cadence (above the background level) in a short time scale, at least in an energy band of 15-25, 25-50, or 50-100 keV. As for the events 
without RHESSI HXR data, we picked up all the events with positive GOES SXR gradients occurring at the same time as DPSs.

%
\subsection{Coincidence of DPS events and HXR/SXR flares}

We find that 49 DPSs events (90\%) with RHESSI are associated with HXR peaks, and 5 events (10\%) are not (Table 2). When considering all the events with RHESSI, HXRS or GOES, 
we found that 95 events (90\%) are correlated with HXR peaks and/or SXR peak gradients, and 11 events (10\%) are not (Table 3). We summarize the results also in pie graphs of Figure 
4, which indicates that most of the DPS events are related to HXR emissions and are associated with particle acceleration. Moreover, we also find that more than 90\% of case (I)-(III) 
events are correlated with HXR peaks, whereas case (IV) shows only 74\% of HXR events. In other words, not all the DPS events are associated with HXR events. 
Especially case (IV) is less correlated with HXR peaks. Considering Neupert effect (1968), it can be rewritten that not all the DPS events are in association with SXR flares.

Actually we confirmed that not all the DPS events are correlated with GOES SXR flares over C-class. Majority of DPSs occurred with HXR and SXR flares, but some DPSs 
with SXR flares and no HXRs (indicating soft flares), some (7 DPS events, attached with $\dagger$ mark in Table 1) with HXRs but without SXRs (because the count of HXRs is so small 
like 40-200 that the SXR component is covered in SXR background), and the others (5 DPS events with $\ast$ mark in Table 1) with neither HXRs nor SXRs (but these were also in 
active regions). For the last 12 events without SXR flares, seven of them are interpreted as independent events from SXR, because the time difference between DPSs and SXR peaks 
is so large. Two of them occurred in decay phase of GOES flares without any enhancement of SXR. One is in the quiet SXR activity. Three of them are neglected because of the small 
increase of SXR like from GOES C5.0 to C5.4. These are probably because the magnetic structure in Case (IV) is complex, which can trap in some cases all accelerated electrons, and 
thus no accelerated electrons reach dense layers of the solar atmosphere, and then much reduced X-ray emission is produced.

Figures 5-7 show typical DPS events associated with HXR bursts. The radiospectra of DPSs are attached with HXR light curves taken by RHESSI, with solid line (15-25 keV), dotted 
line (25-50 keV) and dashed line (50-100 keV). Every event shows an evident emission in the energy band 15-25 keV, but only a small number of events show very weak 
ones in higher energy bands. This indicates that DPSs are more related to particles/plasmas in the energy range of 15-25 keV, rather than in the higher energy range of 25-100 keV. 
The occurrence ratio of different HXRs energy events is summarized in Table 4. All the DPSs associated with HXR peaks taken by RHESSI are accompanied by an enhancement in the 
15-25 keV band. HXRs in the 25-100 keV bands are observed in less than 25\% of all the events which were associated with HXRs in the 15-25 keV range. Case (I) shows the largest 
occurrence ratio of 50-100 keV events to 15-25 keV events, case (IV) show the second, case (II) the third and case (III) shows the smallest.

HXR light curves sometimes show a small hump before the main peak in time, which is well correlated with a DPS event. It is also noted that HXRs in different energy bands show 
time difference between their peaks, i.e. higher energy peaks are reached earlier than lower energy peaks, and DPSs are in general more closely correlated with the higher energy 
peaks in terms of the timing, only when there is a HXR peak in higher energy bands. This can be explained as higher energy particles can precipitate to the chromosphere 
faster than lower energy ones, so the timing of higher energy particles is closer to the timing of particle acceleration in the corona near DPS-emitting plasmoids.

\clearpage

\begin{deluxetable}{llllllllrl}
\tabletypesize{\normalsize}
\tablecaption{The number of events with or without HXR peaks for each case (I)-(IV). HXR data is taken by RHESSI and HXRS. \label{tbl2}}
\tablewidth{0pt}
\startdata
\tableline
\colhead{} & \colhead{case (I)} & \colhead{case (II)} & \colhead{case (III)} & \colhead{case (IV)} & \colhead{Total}\\[+0.1cm]
\tableline
\colhead{with HXR peaks}  & \colhead{5} & \colhead{7} & \colhead{28} & \colhead{9} & \colhead{49}\\[+0.1cm]
\colhead{no HXR peaks} & \colhead{0} & \colhead{0} & \colhead{1} & \colhead{4} & \colhead{5}\\[+0.1cm]
\tableline
\colhead{no RHESSI data} & \colhead{7} & \colhead{4} & \colhead{23} & \colhead{18} & \colhead{52}\\[+0.1cm]
\tableline
\colhead{Total} & \colhead{12} & \colhead{11} & \colhead{52} & \colhead{31} & \colhead{106}\\[+0.1cm]
\enddata
\end{deluxetable}

\begin{deluxetable}{llllllllrl}
\tabletypesize{\normalsize}
\tablecaption{The number of events with or without HXR peaks or GOES SXR gradient peaks for each case (I)-(IV). HXR data is taken by 
RHESSI and HXRS. \label{tbl3}}
\tablewidth{0pt}
\startdata
\tableline
\colhead{} & \colhead{case (I)} & \colhead{case (II)} & \colhead{case (III)} & \colhead{case (IV)} & \colhead{Total}\\[+0.1cm]
\tableline
\colhead{with HXR/SXR grad. peaks}  & \colhead{11} & \colhead{11} & \colhead{50} & \colhead{23} & \colhead{95}\\[+0.1cm]
\colhead{no HXR/SXR grad peaks} & \colhead{1} & \colhead{0} & \colhead{2} & \colhead{8} & \colhead{11}\\[+0.1cm]
\tableline
\colhead{Total} & \colhead{12} & \colhead{11} & \colhead{52} & \colhead{31} & \colhead{106}\\[+0.1cm]
\enddata
\end{deluxetable}

\begin{figure}[btp]
\epsscale{.60}
\plotone{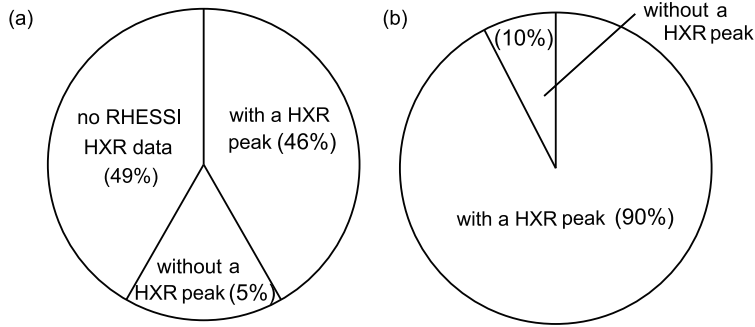}
\caption{(a) The percentage of DPS associated with a HXR peak observed by RHESSI or HXRS. ``Without a HXR peak'' means that there were no HXRs observed 
above the background level. ``No RHESSI HXR data'' means no RHESSI data during DPS events. (b) The same as (a) using the time derivative of the GOES SXR 
flux for ``no RHESSI'' events. \label{fig4}}
\end{figure}

\clearpage

\begin{figure}[hbtp]
\epsscale{.90}
\plotone{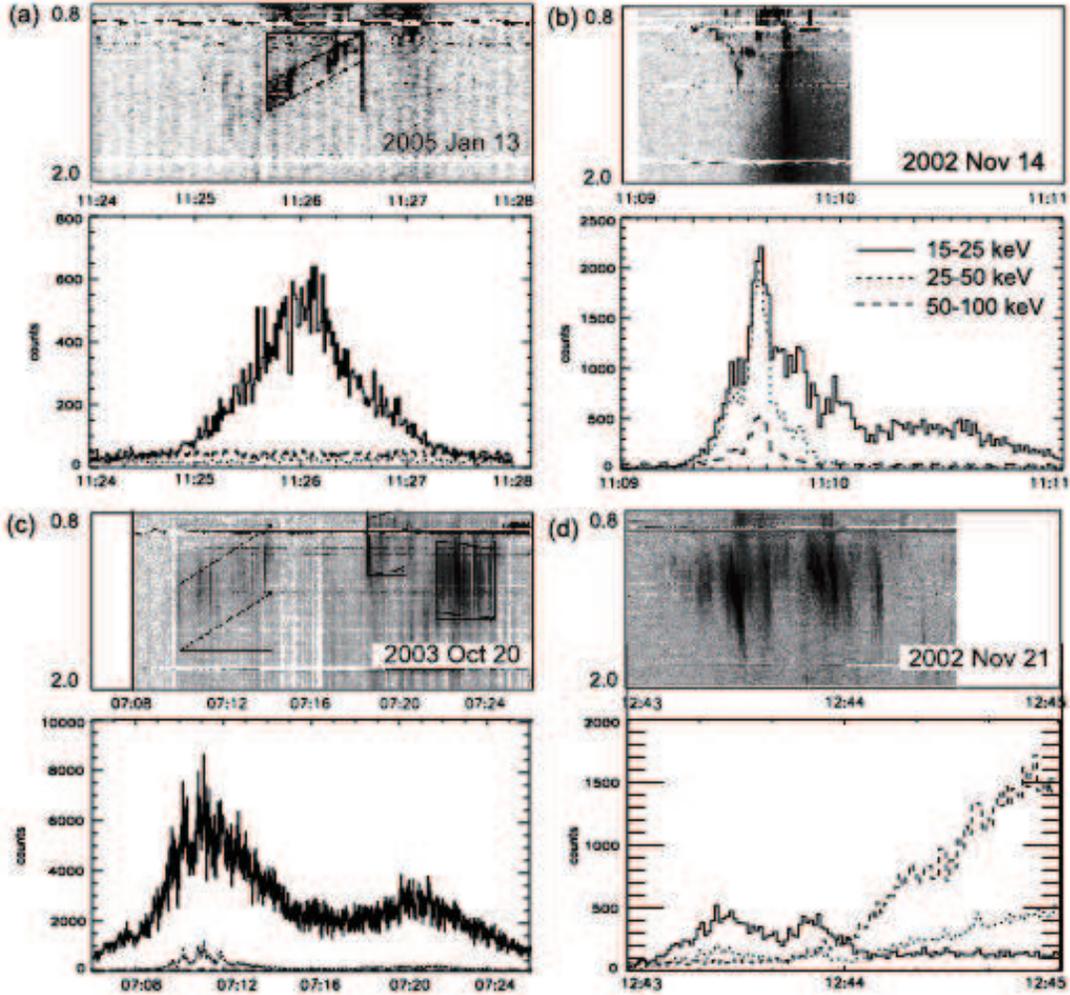}
\caption{Typical DPS events taken by radiospectrograph telescopes in Ond\v{r}ejov observatory and HXR light curves taken by RHESSI in different energy ranges (15-25, 25-50, 
and 50-100 keV). (a)-(d) show events with $\Delta$t$<$0, where $\Delta$t=t$_{DPS}$-t$_{HXR}$. Furthermore, (a) has no HXR$>$25 keV, (b) HXR$<$25 keV coincides with 
HXR$>$25 keV, (c) HXR$<$25 keV coincides with HXR$>$25 keV and HXR peak is long, (d) HXR$<$25 keV precedes HXR$>$25 keV. \label{fig5}}
\end{figure}

\begin{figure}[hbtp]
\epsscale{.90}
\plotone{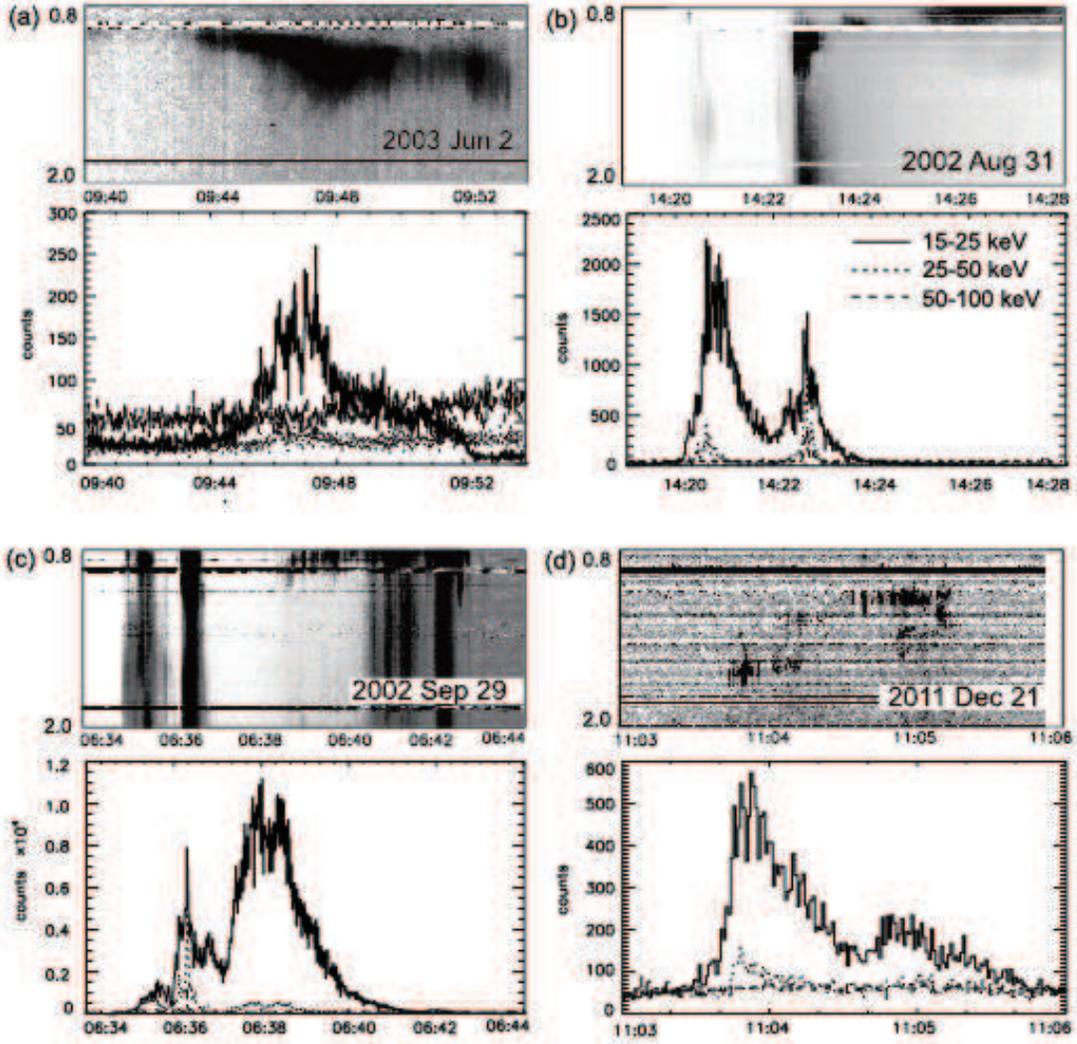}
\caption{Comparison between DPSs observed in Ond\v{r}ejov observatory and HXRs observed by RHESSI in different energy ranges (15-25, 25-50, and 50-100 keV): 
(a) $\Delta$t=0 and no HXR$>$25 keV, (b) $\Delta$t=0 and HXR$<$25 keV coincides with HXR$>$25 keV. (c) $\Delta$t$>$0 and HXR$<$25 keV coincides with HXR$>$25 
keV. (d) $\Delta$t$>$0 and HXR$<$25 keV precedes HXR$>$25 keV. Here $\Delta$t=t$_{DPS}$-t$_{HXR}$. \label{fig6}}
\end{figure}

\begin{figure}[hbtp]
\epsscale{.80}
\plotone{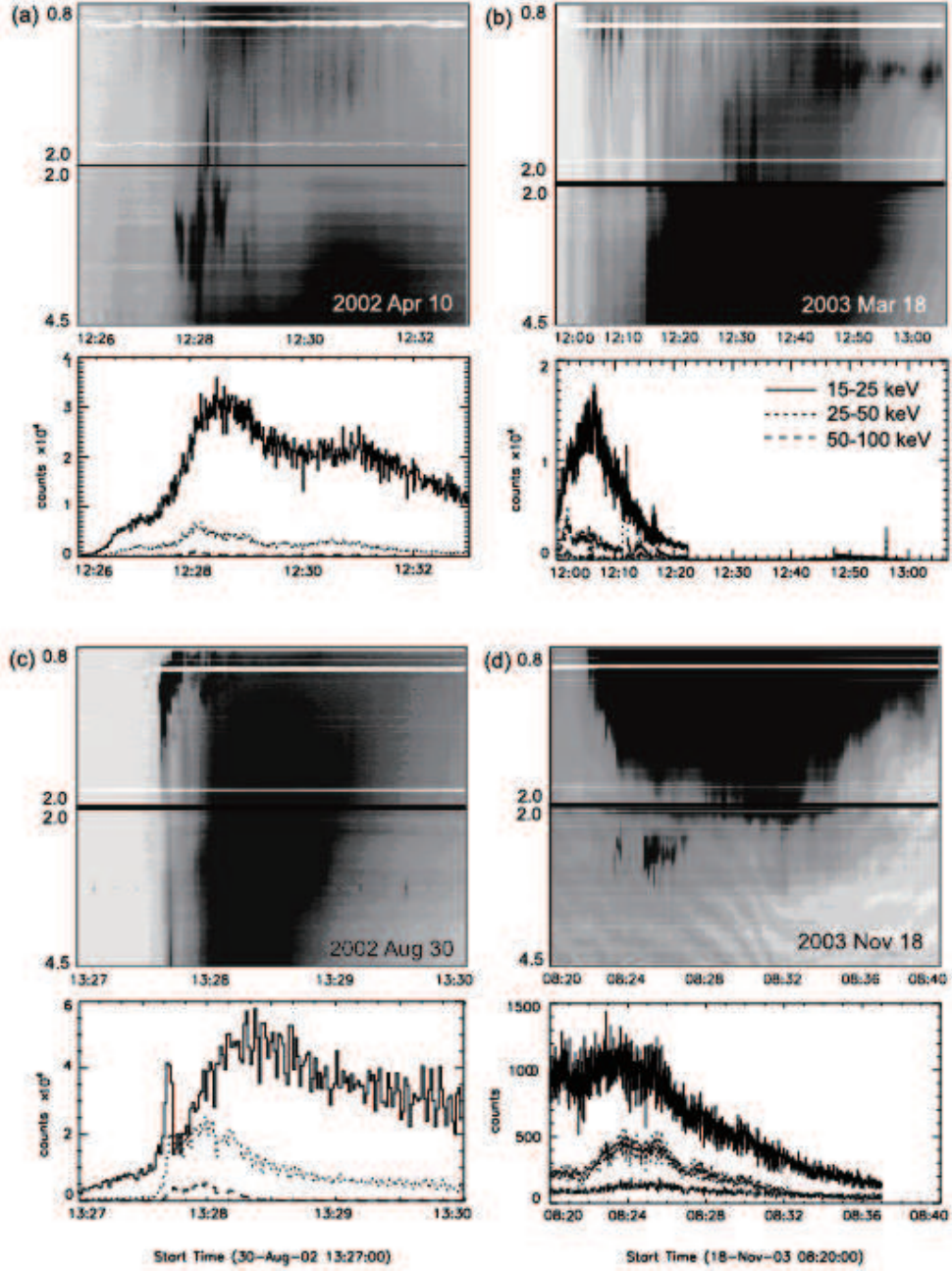}
\caption{DPS events with radio bursts known as broadband continua (or type IV burst) and HXRs observed by RHESSI in different energy ranges (15-25, 25-50, and 50-100 keV): 
(a)-(c) $\Delta$t=0 and $f_{DPS} < f_{burst}$. (d) $\Delta$t$>$0 and $f_{DPS} > f_{burst}$. Here $\Delta$t=t$_{DPS}$-t$_{HXR}$. \label{fig7}}
\end{figure}

\clearpage
%
\subsection{Time difference between DPSs and HXR peaks}

DPS events (with HXR peaks) can be classified by their relation with these peaks. Figure 5 shows events which precedes HXR peaks. Figure 6(a)-6(b) shows DPSs whose acceleration 
or deceleration occur at the same time as HXR peaks, and Figure 6(c)-6(d) shows events delayed to HXR peaks. Here we determine the peak time of HXR (15-25 keV), $t_{HXR}$, 
and the start time of the corresponding DPSs, $t_{DPS}$. The time difference between the two times ($\Delta t$=$t_{DPS}$-$t_{HXR}$) is illustrated in Figure 8. A positive $\Delta t$ 
corresponds to a delay of DPSs relative to HXR, while a negative $\Delta t$ indicates the occurrence of DPSs before the onset or during the rising phase of HXR bursts. In Figure 9, 
most of the events are distributed in the range of -60 s$< \Delta t <$30 s, which can be fitted with a Gaussian function centered at the null value for $\Delta t$, and with its negative-side 
tail enhanced (or double Gaussian functions are also suitable). This means that DPSs mainly occur in the flare ascending phase just after the onset of the flare, and that DPSs occur 
when the gradient of SXRs reaches the maximum \citep[see also][]{wan12}. There are minor extreme events with $|\Delta t| >$60 s, but these may be independent from the HXR activity. 
Furthermore, there is no tendency that the events with larger/smaller $\Delta t$ are associated with larger/shorter duration HXR peaks, i.e. when $\Delta t$ is small 
the events are not always impulsive.

As explained above, the association of DPSs with HXRs can be influenced by the radio emission mechanism and by the trapping in the complex magnetic fields resulting 
from the formation, and possibly the interaction, of multiple plasmoids. The acceleration regions in the current sheet higher in the corona can get magnetically connected and/or 
disconnected multiple times, as the magnetic topology evolves in the layers below \citep[see e.g.][]{bar11}. Namely, new plasmoid formation and coalescence, taking place below the 
acceleration region influences the evolution of the magnetic connectivity of this acceleration region all the way down to the chromosphere/photosphere. Therefore, defining a clear 
correspondence between the radio DPS emissions and HXR remains difficult. Here we propose to look at general tendencies for this connection that can be deduced from observations. 
We interpret cases of positive $\Delta t$ when the radio emission takes more time, for radiation process to take place, e.g. specific velocity distribution, than HXRs. Cases with negative 
$\Delta t$ may indicate that the trapping in a complex magnetic field or complex magnetic connectivity domain leads to particles necessitating more time to precipitate to the chromosphere 
due to the trapping in the magnetic topology, or that very few high energy particles first escaped to the lower dense layers, leading to weak HXRs, while the bulk of high energy particles 
emitting HXRs escaped with a slight delay.
 

The enhancement of HXRs during the acceleration of a plasmoid is only evident for one event. Rather, during multiple plasmoid acceleration events 
or multiple DPS events, initially enhanced HXR emission gradually decreases (28 events, e.g. Fig. 5(c)) or reaches the background level (5 events, e.g. Fig. 6(a)). When 
the HXR curve has multiple enhancements, some DPSs also have multiple enhancements in the intensity of the radio/microwave emission (e.g. Figs. 5(d) and 6(d)).

Figure 7 shows DPS events with additional radio bursts which are called broadband continuum or type IV burst. 6 events out of the 48 DPS events with RHESSI HXR peaks are 
accompanied with radio bursts in the frequency range of 0.8-4.5 GHz. 1 event occurred during the main phase of the radio burst, while 5 events occurred before or at the initial 
phase of the radio burst. It is also noted that one DPS during the main phase is observed in the higher frequency of a radio burst, while 5 DPSs before radio bursts are in the lower 
frequency. Tan et al. (2008) reported that 88\% of DPSs took place at the initial phase of radio bursts (0.6-7.6 GHz) in their analysis. In our analysis, DPSs with radio bursts known 
as the broadband continuum (type IV burst) only represent 13\% of all the cases, and 5/6 (=83\%) of the events occurred at the initial phase of the radio bursts. F\'{a}rn\'{i}k et al. 
(2001) also reported that the GHz broadband radio pulses taken by the 3 GHz Ond\v{r}ejov radiometer are delayed by 2-14 s related to the HXR emission peaks. If DPSs are synchronized 
with HXR peaks, DPSs should be observed 2-14 s prior to the radio bursts.

\begin{figure}[hbtp]
\epsscale{.60}
\plotone{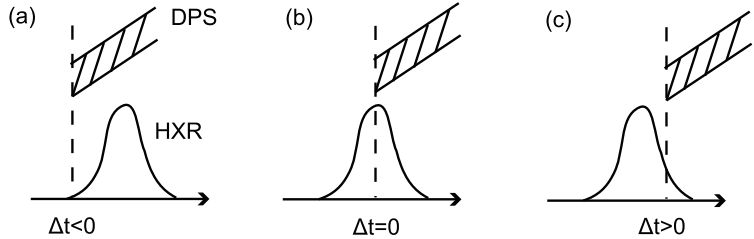}
\caption{The relationship between the start time of a DPS and the peak time of HXRs: (a) DPS precedes HXR, (b) DPS and HXR occur at the same time, (c) DPS is delayed with HXR. Here 
$\Delta$t=t$_{DPS}$-t$_{HXR}$. \label{fig8}}
\end{figure}

\begin{figure}[hbtp]
\epsscale{.80}
\plotone{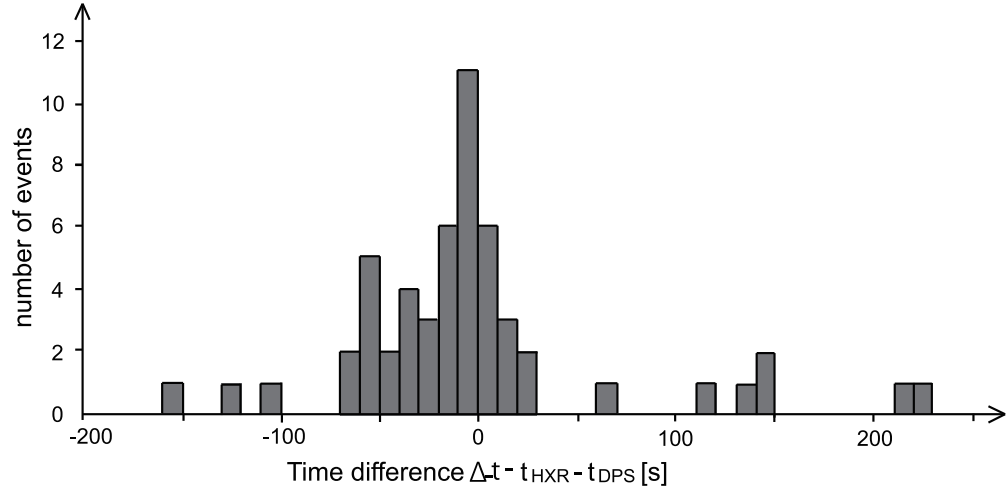}
\caption{Histogram of the time difference between the peak time of HXRs and the start time of an isolated DPS or a series of multiple DPSs. \label{fig9}}
\end{figure}

\begin{deluxetable}{llllrl}
\tabletypesize{\normalsize}
\tablecaption{Distribution of DPS events observed with HXRs from the RHESSI data, for different energy ranges. \label{tbl4}}
\tablewidth{0pt}
\startdata
\tableline
\colhead{} & \colhead{Low Energy} & \colhead{Middle Energy} & \colhead{High Energy}  \\[-0.15cm]
\colhead{} & \colhead{(15-25 keV)} & \colhead{(25-50 keV)} & \colhead{(50-100 keV)} \\
\tableline
\colhead{Case (I)} & \colhead{ 5} & \colhead{ 3}  & \colhead{2} \\ 
\colhead{Case (II)} & \colhead{ 7} & \colhead{ 3} & \colhead{2} \\
\colhead{Case (III)} & \colhead{27} & \colhead{14} & \colhead{5} \\
\colhead{Case (IV)} & \colhead{ 8} & \colhead{ 5} & \colhead{3} \\
\tableline
\colhead{Total} & \colhead{47} & \colhead{25} & \colhead{12} \\ 
\colhead{}        & \colhead{(1.00)} & \colhead{(0.53)} & \colhead{(0.25)} \\ 
\enddata
\end{deluxetable}

\clearpage
%
%
\section{Statistical Results for the Characteristics of Plasmoids}
\subsection{Basic equations, and Derivation of the Plasmoid properties}

It has long been established that fast-drifting radio bursts (type III and reverse-drift bursts) in the decimetric frequency range are caused by radiation from plasma oscillations excited by 
electron beams \citep{wil54}, with typical exciter velocities of $v_b$/c=0.07-0.25 \citep{dul87}. Similarly, in DPSs the fast drifting features were proposed to be generated by the plasma 
emission mechanism with electron beams \citep{kli00}. With this assumption, in the next step, we convert the measured radio frequencies $f_{max}$ and $f_{min}$ into approximated values 
of electron densities, i.e. $n_{max}$ and $n_{min}$, by setting the observed radio frequency equal to the fundamental electron plasma frequency $f_p$, 
\begin{equation}
f_p = 8980 \sqrt{n_e}
\end{equation}
with $n_e$ the electron density (in cgs units). Implicitly it is assumed that the observed decimetric radio emission is related to the fundamental plasma emission. If some bursts were 
emitted at the harmonic level of the plasma frequency, the inferred density would be a factor 4 lower. 

Here we link the atmospheric density with the magnetic field strength, the height of the DPS event, etc. First, the plasma density in the environment surrounding the plasmoid, noted 
hereafter $n_{out}$, is given by models from Alvarez \& Haddock (1973) and Aschwanden \& Benz (1997) \citep[see also Equation 176 in][]{asc02},
\begin{eqnarray}
n_{out}(h)= 
\left\{ \begin{array}{ll}
n_1 \left(\frac{h_1}{h} \right)^p                       & (h \leq h_1),\\
n_Q \exp \left(-\frac{h}{\lambda_T} \right)      & (h \geq h_1)\\
\end{array}\right.
\end{eqnarray}
where $h_1$=1.6$\times$10$^{10}$ cm is the transition height between the lower and the higher corona, $n_1$=4.6$\times$10$^7$, cm$^{-3}$, n$_Q$=4.6$\times$10$^8$ cm$^{-3}$ is 
the base density, $\lambda_T$=$RT/g$=6.9$\times$10$^9$ cm is the thermal spatial scale of density, and the power-law index is $p$=2.38. We can estimate the coronal magnetic field by 
considering the Sun as a magnetic spherical dipole, as suggested in Aschwanden (1999),
\begin{equation}
B_{cor}(h)= B_0 \left(1+\frac{h}{h_D} \right)^{-3} 
\end{equation}
where $B_0$=100 G and $h_D$=7.5$\times$10$^9$ cm. It is noted here that, for the purpose of our study, we need some simple formula approximating the dependence of the magnetic 
field on heights in the corona. In the review book of Aschwanden (2004) (page 19-20), there are two such approximations: the Dulk and McLean's formula which is valid above 1.02$R_{sun}$ 
\citep{dul78} and the Aschwanden's approximation. Both approximations were derived from observations. We used the Aschwanden's approximation, because it is a commonly used approximation 
of the magnetic field in the low corona. It is true that the current layer where the plasmoids are formed is quite a complex region, however, even in simulations of plasmoid formation, the 
ambient magnetic field around the current sheet is often considered to be simple (for example, with a single shearing layer such as often considered Harris sheets). As such, since we are 
mostly interested in the large-scale property of the magnetic field in the region surrounding the plasmoid, taking a simple model for the coronal field such as Aschwanden's one should 
still be a valid approximation.

Furthermore, when the gas pressure inside a plasmoid balances with the gas pressure and the magnetic pressure outside of the plasmoid, the following relationship is satisfied,
\begin{equation}
2n_{in}(h) k_B T_{in} = 2n_{out}(h) k_B T_{out} + \frac{B_{cor}^2(h)}{8\pi} 
\end{equation}
This equation is rewritten as follows,
\begin{equation}
n_{in}(h)= n_{out}(h) \frac{T_{out}}{T_{in}} +\frac{B_{cor}^2(h)}{8\pi}\frac{1}{2k_B T_{in}}
\end{equation}
Here we neglected the term of magnetic pressure, i.e. the guide field, inside a plasmoid for simplicity. If the guide field inside a plasmoid is not negligible, $n_{in}$ is overestimated and gives 
the upper limit. By inserting the equations above (2)-(4) to this pressure balance equation (6), we get the following equation, 
\begin{eqnarray}
n_{in}(h)= 
\left\{ \begin{array}{ll}
n_1 \left(\frac{h_1}{h} \right)^p \frac{T_{out}}{T_{in}}+ \frac{B_0^2}{8\pi 2k_B T_{in}}\left(1+\frac{h}{h_D} \right)^{-6}   &   (h \leq h_1)\\
n_Q \exp \left(-\frac{h}{\lambda_T} \right) \frac{T_{out}}{T_{in}} +\frac{B_0^2}{8\pi 2k_B T_{in}} \left(1+\frac{h}{h_D} \right)^{-6}             &   (h \ge h_1)\\
\end{array}\right.
\end{eqnarray}
Here the temperatures in the corona and surrounding the plasmoid, and in the plasmoid itself, are supposed to be constant. These assumptions can be made since the time scale of the 
thermal conduction $t_{cond}$ is much shorter than the typical time scale $\tau$. Referring to Ohyama \& Shibata (1998) and Nishizuka et al. (2010), we set the coronal temperature 
$T_{cor}$=1.5$\times$10$^6$ K, and the plasmoid temperature $T_{in}$=1.0$\times$10$^7$ K. 

From equation (6), we have obtained a direct relation between the density inside the plasmoid, $n_{in}$, and the height of the DPS emission. By using $f_{max}$ and $f_{min}$, deduced from 
the observations at the start of the DPS event, we can calculate the related $n_{max}$ and $n_{min}$ with equation (2), and deduce the heights $h_{min}$ and $h_{max}$ that give the size 
of the plasmoid $W_{pla}$=$h_{max}$-$h_{min}$ (Fig. 10). Here we assume that the plasmoid is of a similar extent in width as in height. In the following, we 
also define the average height of the plasmoid at the start time $h_{start}$= ($h_{max}$+$h_{min}$)/2 as well as $h_{end}$ for the end of the event. 

Then, the other plasmoid properties can be computed as follows. The plasmoid ejection velocity is given as the difference during $\Delta t$ of the average heights of the plasmoid,
\begin{equation}
v_{pla}= \frac{\Delta h}{\Delta t} = \frac{h_{end}-h_{start}}{t_{end}-t_{start}}
\end{equation}
The Alfv\'{e}n speed and the plasma $\beta$, the ratio between thermal and magnetic pressures, are calculated from the outside coronal magnetic field and density,
\begin{eqnarray}
v_A(h)=\frac{B_{cor}(h)}{4\pi m_p n_{out}(h)}\\
\beta = \frac{2n_{out}(h)k_B T_{out}}{B_{cor}^2(h)/8\pi}
\end{eqnarray}
Assuming an incompressible plasma in the reconnection region (such as Sweet-Parker reconnection model), the reconnection rate $M_A$ is expressed by the mass conservation equation 
between the inflow and the outflow: $v_{in} L_{in}$= $v_{pla} W_{pla}$ \citep[equation (1); see also][]{shi01}. Then, the reconnection rate, calculated with the normalized inflow velocity, as well 
as the energy release rate, are given as follows,
\begin{eqnarray}
M_A = \frac{v_{in}}{v_A} = \frac{W_{pla}}{L_{in}} \cdot \frac{v_{pla}}{v_A}\\
\frac{dE}{dt} = 2\frac{B_{cor}(h)^2}{4\pi} v_{in}L_{in}^2
\end{eqnarray}

In the following, we assume that $L_{in}$ is similar to the height of the DPS emission at the start of the event, $h_{start}$ (Fig. 10). This is acceptable, when the length of a current sheet 
is much larger than the height of the loop-top ($L_{in}$=$h_{start}$-$h_{loop-top}$-$\frac{W_{pla}}{2}$ $\sim h_{start}$). On the basis of our calculation result (Fig. 11 in the 
next section), it is true that the estimated height of the plasmoid is (2-4)$\times$10$^9$ cm, which is larger than the typical loop-height ($\sim$10$^9$ 
cm). Some EUV observations of plasmoid ejection show that the current sheet can extend much further above the flare loops \citep[e.g.][]{lin00, web03, sav10, ree11}. 
However, in cases considered here with DPS events associated with HXR peaks and therefore occurring in the impulsive phase at the start of the flare, the current sheet is expected much 
shorter than several $R_{sun}$ what Savage et al. (2010) find several hours after the loss of equilibrium. At the time of the DPS, the flare loops could account for 50\% of the calculated 
plasmoid height. $L_{in}$ also includes the added radius of the plasmoid at the top of the current sheet. The uncertainties of this $L_{in}$ propagate through to the determination of the 
reconnection rate $M_A$ in the following equation;
\begin{equation}
\frac{|\Delta M_A|}{M_A} = \frac{|\Delta W_{pla}|}{W_{pla}} + \frac{|\Delta L_{in}|}{L_{in}} + \frac{|\Delta v_{pla}|}{v_{pla}} + \frac{|\Delta v_A|}{v_A}
\end{equation}
Therefore, the uncertainty of $M_A$ is also 50-90\% from the determination and measurement errors of $L_{in}$, in addition to the errors from the other parameters.

\begin{figure}[hbtp]
\epsscale{.50}
\plotone{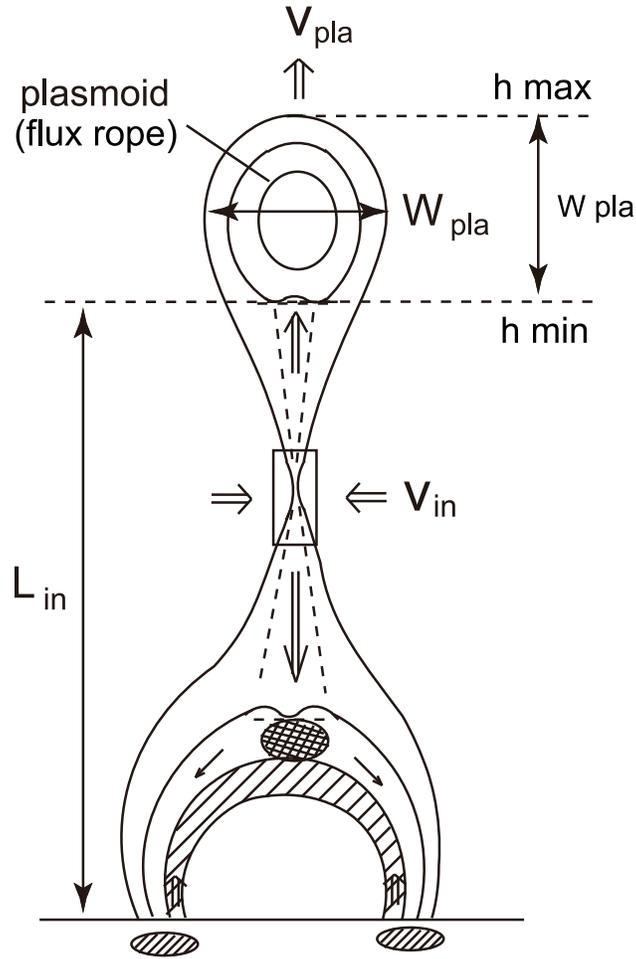}
\caption{Cartoon of a flare eruption with the plasmoid ejection velocity $v_{pla}$, the width of a plasmoid $W_{pla}$, the length of the inflow region $L_{in}$ and the inflow 
velocity $v_{in}$.\label{fig10}}
\end{figure}

\clearpage
%
%
\subsection{Distributions of Velocity, Width, Height of Plasmoids and Reconnection Rate: Case {\bf (I)+(II)} and {\bf (III)}}

Figure 11 shows histograms of the distribution of the velocity, the width, the height of ejected plasmoid and the reconnection rate, which are calculated from the DPS events of cases 
(I)+(II) and (III). We consider cases (I) and (II) together as they have a similar definition (events with constant frequency drifts) and as to allow direct comparison with the more numerous 
DPSs of case (III). We also do not show the histograms for case (IV)-DPSs, as these events are rather complicated with multiple events with non-constant frequency drifts happening 
at the same time: the histogram would be biased by how we would separate each individual event.

Interpreting DPSs as signatures of plasmoids, most of the cases (I)+(II)-events show upward motions of plasmoids, while the case (III)-events contain both upward and downward plasmoid 
events. The case (III)-events show acceleration or deceleration of plasmoids and velocity change in time. The two phases before and after acceleration/deceleration are counted twice 
in case (III) and events with n-times velocity changes are counted (n+1) times, similar to B\'{a}rta et al. (2008). The velocity of plasmoids is mainly in the range of 0-5$\times$
10$^7$ cm s$^{-1}$ (Fig. 11(a)). The 60-70\% of case (III) are upward events, while 30-40\% are downward events. The distributions of upward and downward events are almost the same. 
The width of a plasmoid is typically (0.3-1.1)$\times$10$^9$ cm, and the height of the plasmoid, i.e. the length of a current sheet attached below the plasmoid, is typically (1.8-3.6)$\times$
10$^9$ cm (see Figs. 11(b)-11(c)). This is the first paper to estimate the width of the plasmoid and the length of a current sheet (W$_{pla}$ and L$_{in}$) from DPS events, and the original 
histograms of the instantaneous bandwidth and the starting frequency are in almost the same range of B\'{a}rta et al. (2008). The typical duration of DPSs is 0-60 s (Fig. 11(d)). The 
reconnection rate derived from the DPS events of cases (I)+(II) and (III) is typically less than 0.04 (Fig. 11(e) and 11(f)). This is comparable to the ones found in previous studies \citep[M$_A$
=0.001-0.1, e.g.][]{tsu97, ohy97, yok01, iso02, asa04, nar06, tak12}.

Then what determines and/or controls the reconnection rate? The dependence of the reconnection rate on the plasmoid velocity, the width (size), the plasma beta, and the aspect 
ratio of a current sheet attached below a plasmoid is shown in Figures 12(a)-12(e) for cases (I)+(II) and (III). The reconnection rate has a good correlation with the plasmoid velocity. 
This tendency is retained even after normalizing the plasmoid velocity by the Alfv\'{e}n velocity. On the other hand, the width and the aspect ratio of plasmoids are weakly correlated 
with the reconnection rate (Figs. 12(c) and 12(d)). The plasma beta is mainly distributed in the low beta regime (0.02-0.04), and it seems that the distribution of the reconnection rate 
is broader with smaller plasma beta. Furthermore, the relationship between the ejection velocity and the width of plasmoids is shown in Figure 12(f). For larger plasmoid with W$_{pla} >$ 
10$^9$ cm, the plasmoid velocity tends to become smaller as the size becomes larger. This is probably because the larger plasmoids are heavier and are consequently 
more decelerated by the gravity force than smaller plasmoids. This result is consistent with B\'{a}rta et al. (2008). On the other hand, for smaller plasmoids with W$_{pla}<$6
$\times$10$^8$ cm, the plasmoid velocity tends to increase as the width of the plasmoid increases. This may indicate that larger plasmoid induces stronger inflow, enhancing the 
reconnection rate and accelerating itself by the faster reconnection outflow.

\begin{figure}[hbtp]
\epsscale{.90}
\plotone{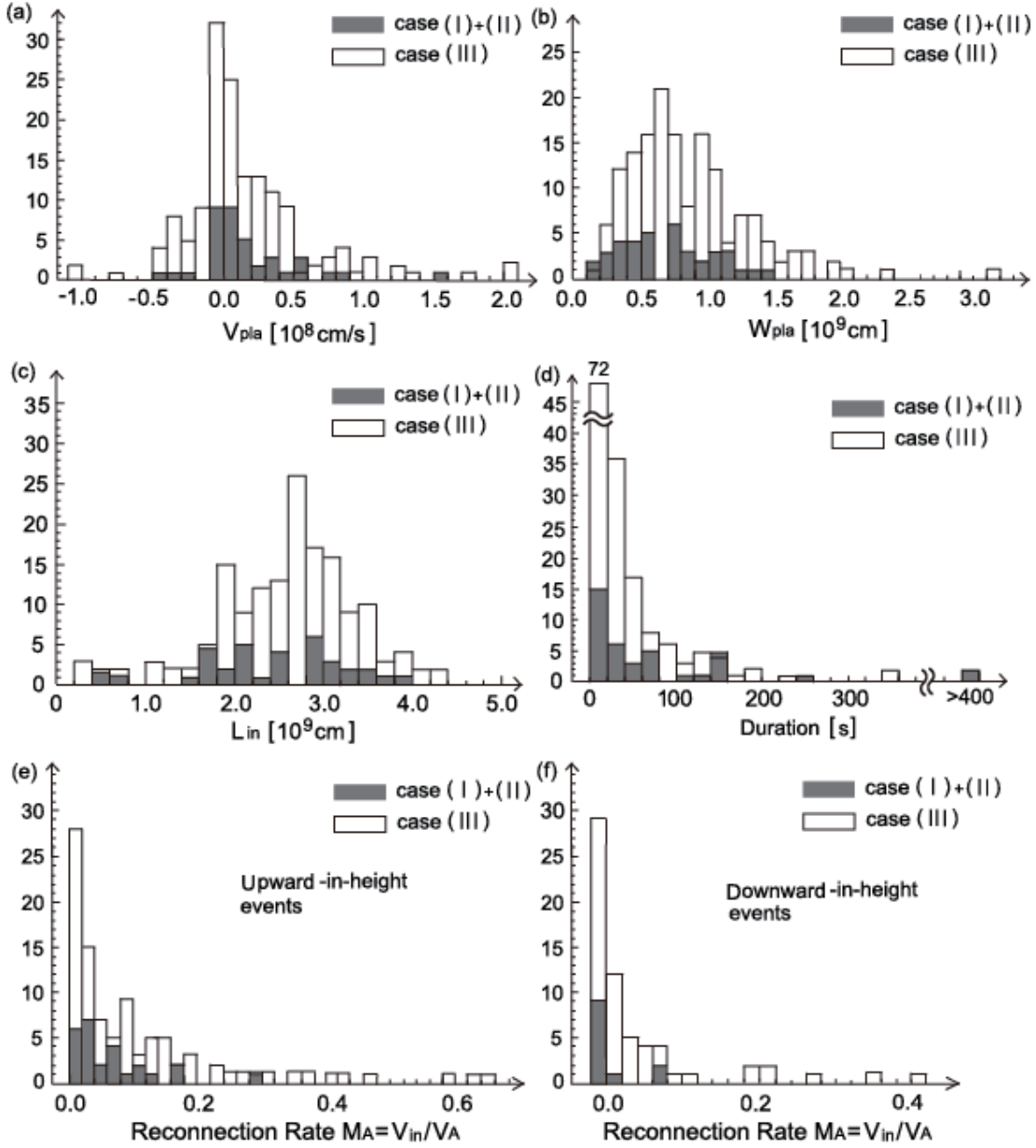}
\caption{Histograms of the distributions of (a) the ejection velocity, (b) the width and (c) the height of plasmoids, (d) the duration and (e)-(f) the reconnection rate during the upward/
downward-in-height plasmoid ejections, which are derived from the DPS events of cases (I)+(II) and (III) during 2002-2012. \label{fig11}}
\end{figure}

\begin{figure}[hbtp]
\epsscale{.75}
\plotone{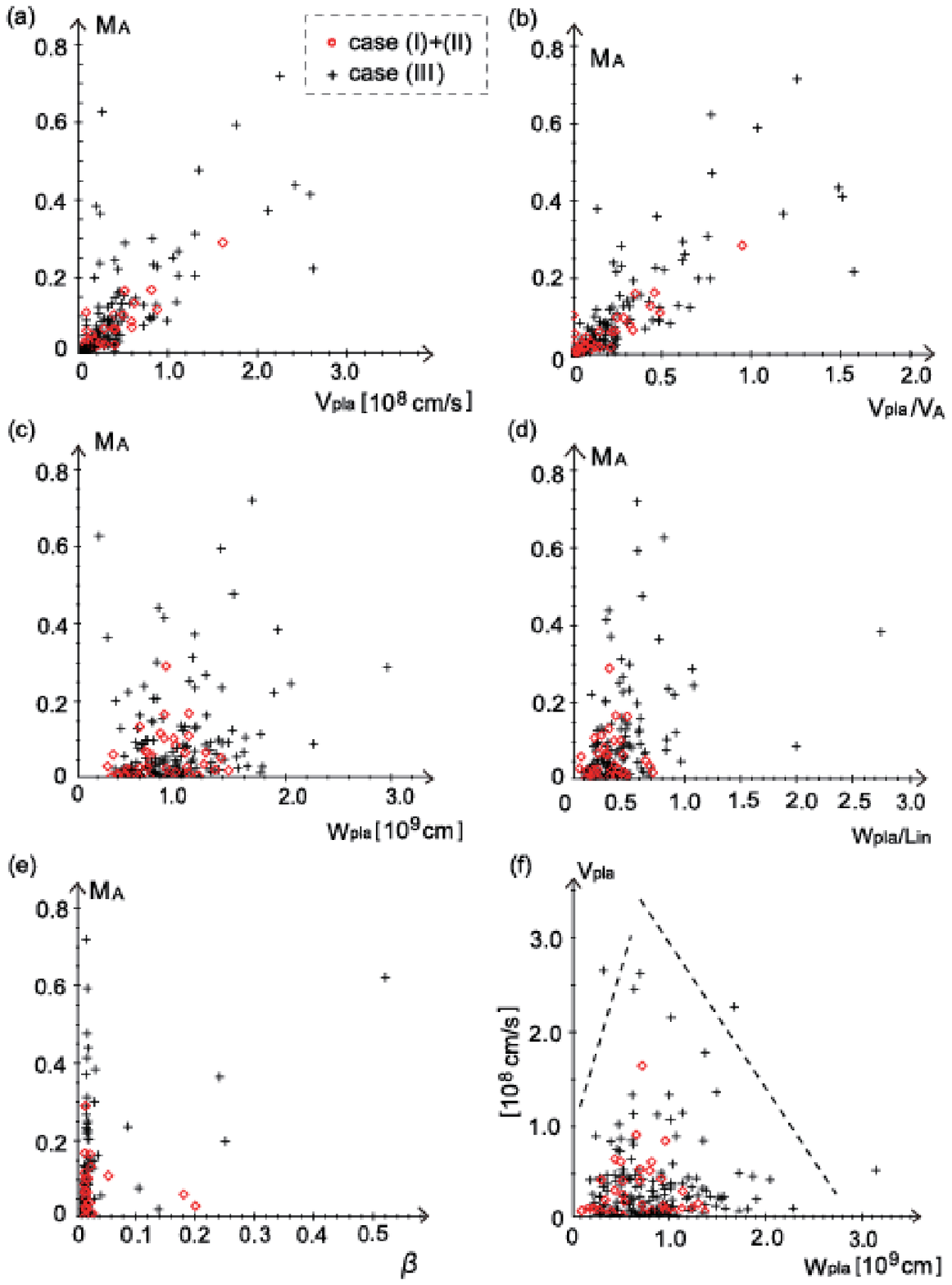}
\caption{(a)-(e) Correlation study for cases (I)+(II) and (III), showing the dependence of the reconnection rate on the original plasmoid velocity, the plasmoid velocity normalized by the 
local Alfv\'{e}n velocity, the plasmoid width, the aspect ratio of the reconnection region (i.e. the ratio of the width and the length of a current sheet formed below a plasmoid) and the 
plasma beta. (f) The relationship between the velocity and the width of plasmoids. \label{fig12}}
\end{figure}

%
%
\subsection{Correspondence with EUV and X-ray Plasmoid ejection}

Next we compared DPS events with imaging observations in EUV emissions taken by the Atmospheric Imaging Assembly (AIA) on board SDO, to identify a possible correlation between 
DPSs and plasmoids seen in the EUV corona. Since the temporal resolution of AIA is limited to 12 s, only long-duration DPS events ($>$1 minutes) observed after SDO launch are used 
for analysis. 

Figure 13(a)-(c) shows snapshot images of a flare event on 2012 June 13 taken with the AIA 94 \,\AA\ filter. The event shows a plasma ejection in the south-east direction between 
12:32-13:17 UT. The loops are gradually moving outward around 12:32 UT and the motion accelerates at 12:53 UT up to an apparent velocity of 2$\times$10$^6$ cm 
s$^{-1}$. In this event, there are no RHESSI data, but the steepest slope of the GOES SXR flux was observed at the beginning of the outward motion. In the 211 \,\AA\ images small 
plasma blobs are seen to move in the trailing region of the outward-moving loops. We found a downward motion blob during 12:50:33-12:51:09 UT, and then two small plasma blob ejections 
between 12:51:57-12:52:33 UT. The radio spectrograph data during the same period is shown in Figure 13(b). There are three periods of DPS: a downward-in-height (upward-in-frequency) 
DPS (12:50:00 UT), a constant frequency DPS (12:50:40 UT), and an upward-in-height (downward-in-frequency) DPS (12:52:30 UT). The upward motion of the DPS after 12:52:30 UT 
corresponds to the timing of the outward moving coronal loops observed in 94 \,\AA\, and the upward velocity of the DPS is 2$\times$10$^7$ cm s$^{-1}$, which is 
much larger than the apparent velocity (2$\times$10$^6$ cm s$^{-1}$). The downward motion of the DPS after 12:50 UT has no clear corresponding features in 94 
\,\AA\ images, but small scale blobs in 211 \,\AA\ images may be candidates to explain the observed DPSs. 

Figure 14(a) shows a flare event on 2012 May 8, which shows footpoint brightenings followed by multiple plasma ejections and a loop expansion: the initial ejection at 13:05:26 UT, the 
second one at 13:06:02 UT, the loop expansion during 13:07-13:08 UT and the third ejection at 13:08:02 UT. The radio spectrum of DPSs is shown in Figure 14(i), in which three DPSs 
are observed: a slow rising-in-height DPS (13:04:50 UT), a downward-in-height DPS (13:07:00 UT) and another slow rising-in-height DPS (13:07:20 UT). The first upward-in-height DPS 
occurred at the beginning of the first and the second eruptions in 94 \,\AA\ and 193 \,\AA\ images, the second downward-in-height DPS is during the loop expansion above the active 
region but there seems to be no clear corresponding moving feature, and the third DPS is at the beginning of the double plasma ejection.

Figure 15(a) shows an east-limb flare event on 2012 July 6. In 304 \,\AA\ images, a cool prominence upward motion (marked by an arrow) was observed around 13:28-13:30 UT, which was 
located below a cavity observed in 94 \,\AA\ images and at the footpoint of the dark open field lines in the 131 \,\AA\ images. In the 304 \,\AA\ snapshot images, small scale intermittent 
plasma movements are also observed inside the loop structure, apart from the previous prominence. On the other hand, an upward(-in-height) DPS occurred at 13:29:22 UT (Fig. 15(b)), 
when the cool prominence in 304 \,\AA\ showed upward motion and when the slope of the GOES SXR flux became the steepest. The apparent upward velocity is 1$\times$
10$^6$ cm s$^{-1}$ which is the vertical velocity on the limb, while the upward velocity estimated from the DPS is 1$\times$10$^7$ cm s$^{-1}$. The later radio burst 
after 13:30:25 UT may be related to the inner plasma flow inside the coronal loop structure. 

Here we note that, in the paper by B\'{a}rta et al. (2008) and also in the present study, we found that the DPS with larger instantaneous bandwidth usually drifts slower than that with 
smaller bandwidth. It means that larger plasmoids move slower than smaller ones. It is highly probable that in the current sheet there are several plasmoids with different sizes. In EUV 
we recognize only the largest plasmoid and this EUV plasmoid need not be the same as the plasmoid, where the DPS is generated. It may explain the difference in the velocity of the EUV 
plasmoid and that derived from the DPS on 2012 July 6. It is true that the size (width) of the observed EUV plasmoids is in the range of 5''-40'', i.e. (0.4-3.6)$\times$10$^9$ cm. This is 
comparable to but a little larger than the typical width of the DPS plasmoids (0.3-1.1)$\times$10$^9$ cm. 

%
 
\begin{figure}[hbtp]
\epsscale{.75}
\plotone{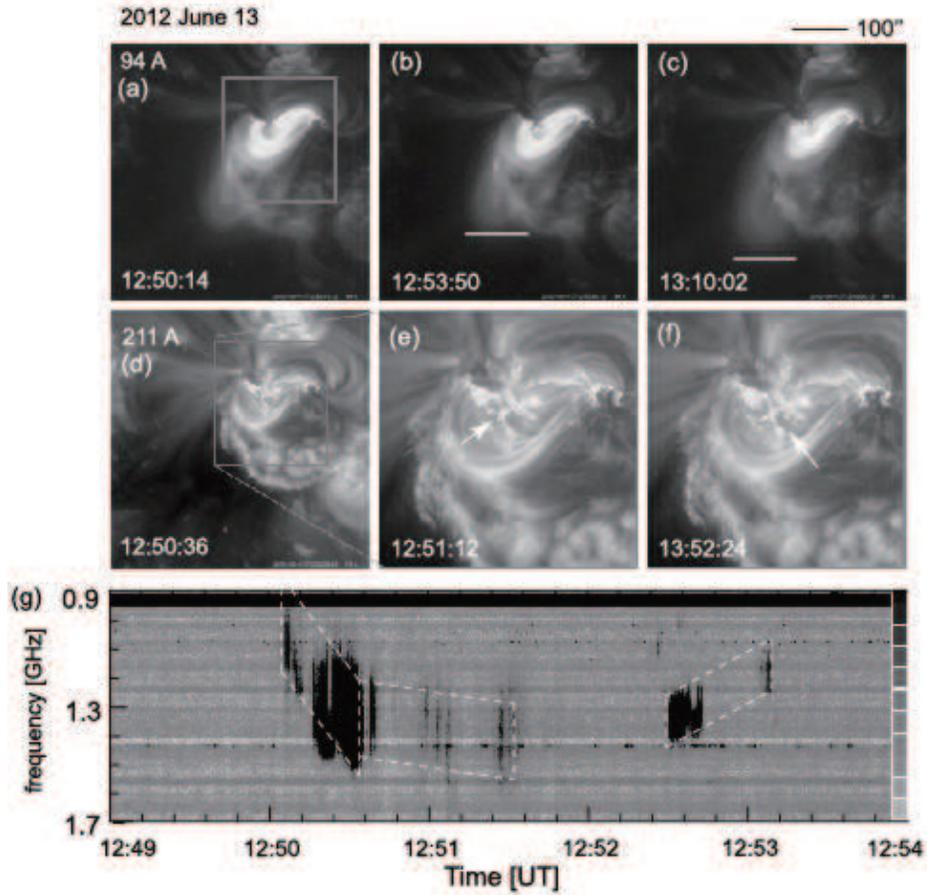}
\caption{(a)-(f) Snapshot images of a flare event on 2012 June 13, taken by 94 and 211 \,\AA\ filters of AIA/SDO. In (b) and (c), the front of a plasma ejecta is shown by a solid line. 
(e)(f) are zoom-up images of (d) taken by 211 \,\AA\, whose field-of-view is shown by a square line in (d). Arrows show small plasma blobs moving in the coronal loops. (g) A radio spectrum 
of DPSs on 2012 June 13, taken by the radiospectrograph telescope in Ond\v{r}ejov observatroy. \label{fig13}}
\end{figure}

\begin{figure}[hbtp]
\epsscale{.90}
\plotone{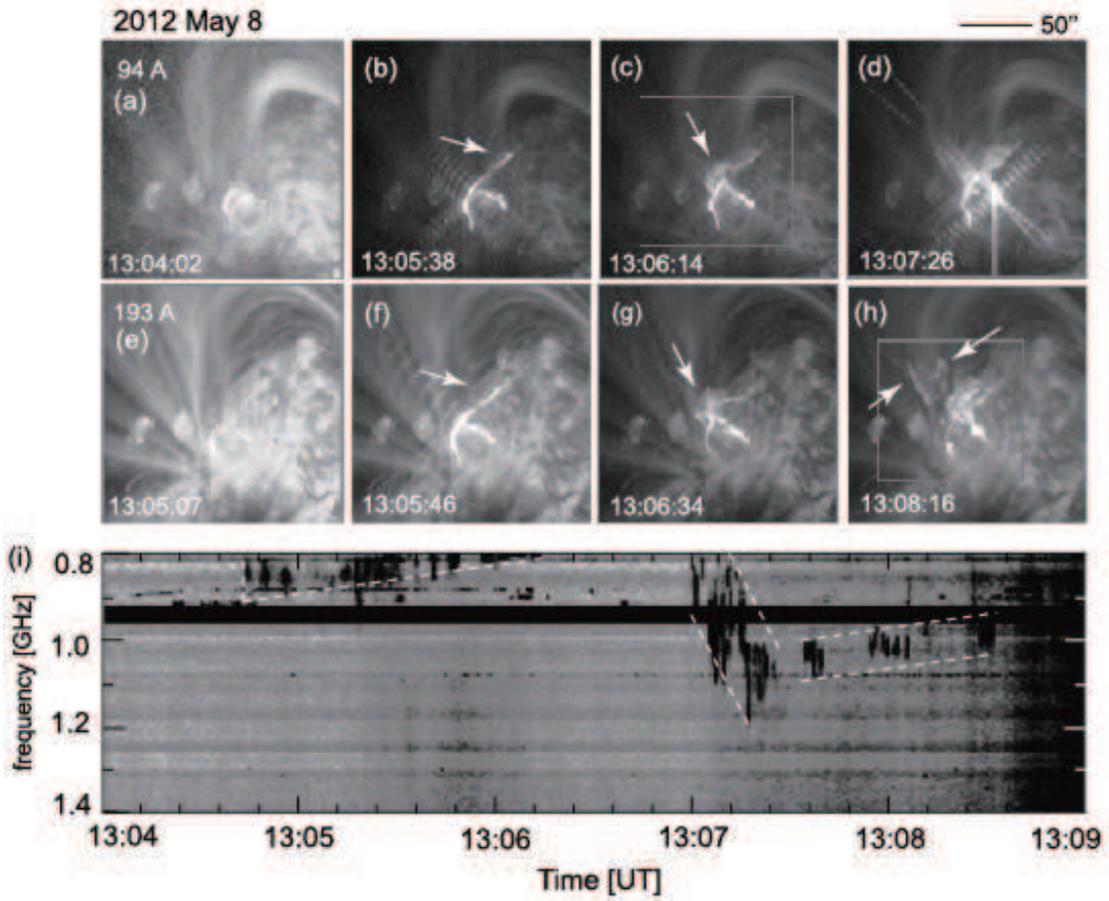}
\caption{(a)-(h) Snapshot images of a flare event on 2012 May 8, taken by 94 and 193 \,\AA\ filters of AIA/SDO. The event shows two ribbon structure and plasma ejections marked by 
arrows. (i) A spectrum of DPSs on 2012 May 8, taken by the radiospectrograph telescope in Ond\v{r}ejov observatory. \label{fig14}}
\end{figure}

\begin{figure}[hbtp]
\epsscale{.85}
\plotone{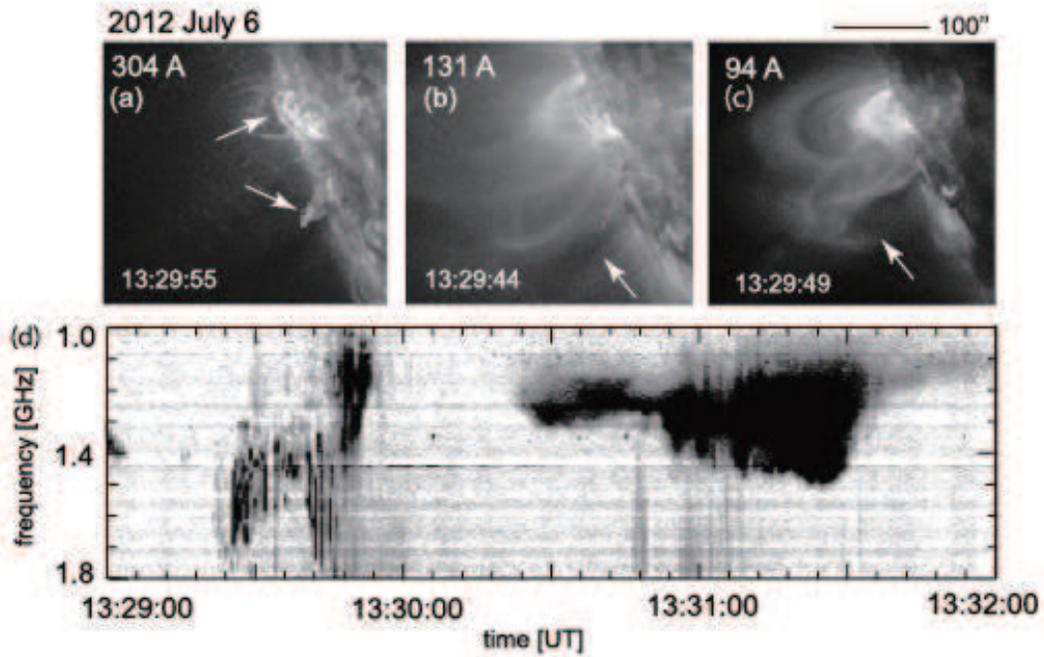}
\caption{(a)-(c) Snapshot images of a flare event on 2012 July 6, taken by 304, 131 and 94 \,\AA\ filters of AIA/SDO at almost the same time. Arrows in (a) show moving cool plasma in 
coronal loops (above) and an eruption of a cool plasma (below). In (b)(c), a cavity is shown by an arrow, which is located at the footpoint of the cool plasma ejection observed in 304 \,\AA\ 
images. \label{fig15}}
\end{figure}

\clearpage
%
\section{Summary and Discussion}

We statistically analyzed 106 DPS events in the frequency range of 0.8-4.5 GHz, observed with radiospectrographs in Ond\v{r}ejov observatory between 2002 and 2012. The number of 
DPSs increases with the solar activity, and it peaked at around 2002 and 2012. The DPSs were classified into 4 groups: (I) single events with a constant frequency drift [12 events], (II) 
multiple events occurring in the same flare with constant frequency drifts [11 events], (III) single or multiple events with increasing or decreasing frequency drift rates [52 events], and 
(IV) complex events containing multiple events occurring at the same time in the different frequency range [31 events]. Most of the DPS events show an increase or a decrease in their 
frequency drifts. 70\% of DPSs drifted toward lower frequency, while 30\% drifted toward larger frequency. In other words, interpreting DPSs as plasmoids, most plasmoid ejections are 
accelerated or decelerated, and 70\% of plasmoids are ejected upward and 30\% downward. 

Of the DPS events with simultaneous RHESSI observations, 90\% of events occurred in association with HXR bursts, mainly in the energy range of 15-25 keV, and they 
occurred at the beginning or at the peak time of HXR burst. 90\% of DPSs were associated with the peak of the GOES SXR gradient and/or RHESSI HXR burst. With the 
exception of the large flares, HXR photons with energy $\>$25 keV is typically produced by nonthermal electrons, whereas the origin for the HXR in the energy range of 15-25 keV is a 
mixture of thermal and nonthermal electrons. This is why our result indicates that most of the events are associated with particle acceleration and partially plasma heating. 5 events 
were observed with radio bursts known as the broadband continuum (type IV burst), and they occurred at the beginning of the radio bursts in the higher frequency range, except for one 
event. When the duration of the HXR burst was long, multiple DPSs occurred. 
 
Interpreting DPSs as signatures of plasmoids, we measured the ejection velocity, the width and the height of a plasmoid, from which we estimated the inflow velocity and the reconnection 
rate by considering the mass conservation of incompressible plasma: the velocity 0-5$\times$10$^7$ cm s$^{-1}$, the width (0.3-1.1)$\times$10$^9$ cm, the height (= the 
current sheet length) (1.8-3.6)$\times$10$^9$ cm, the duration 0-60 s, and the reconnection rate 0-0.04. The width is smaller than the height of the plasmoid, but they are comparable 
to each other (consistent with B\'{a}rta et al. (2008)). This means that DPS plasmoids are observed in the lower corona, just after the formation and the ejection. This is different from 
other events of larger X-ray plasmoid ejection \citep[e.g.][]{gle13}. There is a positive correlation between the reconnection rate and the ejection velocity, while the aspect ratio 
of a current sheet has no clear correlation with the reconnection rate. For larger plasmoid (W$_{pla}>$10$^9$ cm), the plasmoid velocity tends to decrease as the size 
becomes larger, whereas for smaller plasmoids (W$_{pla}<$6$\times$10$^8$ cm), the plasmoid velocity tends to increase as the width of the plasmoid increases. This is 
probably because the larger plasmoids are heavier and are consequently more decelerated by the gravity force than smaller plasmoids. When plasmoids are small, the gravity force is too 
small to affect the plasmoid dynamics, rather the plasmoid velocity increases as the width of the plasmoid increases, probably because larger plasmoids induces more inflow, enhancing 
the reconnection rate and additionally accelerates themselves  by fasten reconnection outflow.

Some of the DPSs show plasmoid counterparts in AIA images, i.e. plasma ejections or blobs, but others are not clear. This is probably because the emission mechanisms of DPSs and EUVs 
are different and because the time scale of DPSs is too short compared with the AIA time cadence. DPSs are always associated with HXR in 15-25 keV, but not always in 25-100 keV. 
Power-law indices of HXR spectra are soft. This indicates that the DPS emission mechanism is not determined by the energy of particles but rather by the condition of the bump-in-tail 
distribution in electron velocity ($dF(v)/dv>$0 for $v>$0; $F(v)$ being the velocity distribution function), which produces the instability for which a wave-particle interaction occurs.

In laboratory experiments, the reconnection electric field is enhanced when a plasmoid is ejected out of a current sheet \citep{ono11, hay12}. This enables a more efficient particle 
acceleration and a reconnection rate enhancement, which is consistent with our results. It is also noted that a larger plasmoid ejection leads to a larger enhancement of the electric 
field and thus a larger reconnection rate in laboratory experiments. However, in our analysis as well as that of B\'{a}rta et al. (2008), smaller plasmoids induce a larger reconnection rate. 
This is probably because larger plasmoids in the solar atmosphere are decelerated by the gravity force, which does not play a strong effect in a laboratory confined plasma configuration. 
This can explain the differences between laboratory experiments and the solar atmosphere.

The intensity of HXR emission can be described by using Neupert effect as follows; 
\begin{equation}
I_{HXR} \propto \frac{dI_{SXR}}{dt} \propto \frac{dEth}{dt} \sim \frac{dEkin}{dt} = \frac{d}{dt}\left( \frac{mv_{pl}^2}{2}\right) = mv_{pl} \cdot \frac{dv_{pl}}{dt}
\end{equation}
Here we assumed equipartition of released energy to thermal and kinetic energies. This indicates a positive correlation between the HXR intensity and the plasmoid velocity and acceleration. 
Since the HXR intensity is empirically believed to be proportional to the reconnection rate \citep{asa04}, the plasmoid velocity is also proportional to the reconnection rate, as well as 
equation (1) proposed in plasmoid-induced reconnection model \citep{shi01}. 

We classified 106 DPSs into 4 cases and mainly analyzed cases (I)-(III). Case (IV) shows multiple DPSs at the same time individually moving upward and downward. Some DPSs in case (IV) 
show merging and separation processes, which may indicate that the coalescence and split of plasmoids take place. With recent numerical simulations, multiple plasmoid forming at different 
scales in a current sheet and particle accelerations in association with plasmoid dynamics have been intensively studied. The comparison between the events occurring in the corona 
and simulations may give us further understandings of energy release by magnetic reconnection and particle acceleration, and the onset of flares, eruptions and radio bursts.

\acknowledgments
We first acknowledge an anonymous referee for his/her useful comments and suggestions. This work was supported in part by JSPS KAKENHI Grant Number 24740132
and in part by the JSPS Core-to-Core Program 22001. MK and MB acknowledge the support by grant P209/12/0103 (GA CR).\\

%
%

\clearpage

%
%
%

\begin{thebibliography}{}
%
\bibitem[Alvarez \& Haddock (1973a)]{alv73a} Alvarez, H. \& Haddock, F. T., 1973, \solphys, 30, 175
%
%
%
\bibitem[Asai et al. (2004)]{asa04} Asai, A., Yokoyama, T., Shimojo, M., Masuda, S., Kurokawa, H. \& Shibata, K., 2004, \apj, 611, 557
%
\bibitem[Aschwanden \& Benz (1997)]{asc97} Aschwanden, M. J. \& Benz, A. O., 1997, \apj, 480, 825 
%
\bibitem[Aschwanden et al. (1999)]{asc99} Aschwanden, M. J., Newmark, J. S., Delaboudini\'{e}re, J.-P., et al., 1999, \apj, 515, 842
%
\bibitem[Aschwanden (2002)]{asc02} Aschwanden, M. J., 2002, Space Sci. Rev., 101, 1 
%
\bibitem[Aschwanden (2004)]{asc04} Aschwanden, M. J., 2004, Physics of the solar corona, Springer, Praxis Publ. Chichester, UK 
%
\bibitem[Bain et al. (2012)]{bai12} Bain, H. M. , Krucker, S., Glesener, L. \& Lin, R. P., \apj, 2012, 750, 44
%
\bibitem[B\'{a}rta et al. (2008)]{bar08} B\'{a}rta, M., Karlick\'{y}, M \& \v{Z}emli\v{c}ka, R., 2008, \solphys, 253, 173
%
\bibitem[B\'{a}rta et al. (2012)]{bar12} B\'{a}rta, M., Sk\'{a}la, J., Karlick\'{y}, M., \& B\"{u}chner, J., 2012, Multi-scale Dynamical Processes in 
Space and Astrophysical Plasmas, Astrophysics and Space Science Proc., 33, 43 (Springer-Verlag Berlin Heidelberg 2012)
%
\bibitem[B\'{a}rta et al. (2011)]{bar11} B\'{a}rta, M., B\"{u}chner, J., Karlick\'{y}, M., \& Sk\'{a}la, J., 2011, \apj, 737, 24  
%
%
\bibitem[Benz et al. (2005)]{ben05} Benz, A. O., Grigis, P. C., Csillaghy, A., Saint-Hilaire, P., 2005, \solphys, 226, 121
%
%
%
%
%
%
%
\bibitem[Drake et al. (2006)]{dra06} Drake, J. F., Swisdak, M., Che, H. \& Shay, M. A., 2006, Nature, 443, 553
%
\bibitem[Dulk \& McLean (1978)]{dul78} Dulk, G. A. \& McLean, D. J., 1978, \solphys, 57, 279
%
\bibitem[Dulk et al. (1987)]{dul87} Dulk, G. A., Goldman, M. V., Steinberg, J. L., \& Hoang, S., 1987, A\&A, 173, 366
%
\bibitem[F\'{a}rn\'{i}k et al. (2001)]{far01} F\'{a}rn\'{i}k, F., Garcia, H., \& Karlick\'{y}, M., 2001, \solphys, 201, 357
%
%
%
\bibitem[Glesener et al. (2013)]{gle13} Glesener, L., Krucker, S., Bain, H. M., Lin, R. P., 2013, \apj, 779, 29
%
\bibitem[Hayashi et al. (2012)]{hay12} Hayashi, Y., Ii, T., Inomoto, M., \& Ono, Y., 2012, IEEJ Transactions on Fundamentals and Materials, 132, 239
%
\bibitem[Hoshino (2012)]{hos12} Hoshino, M., 2012, PRL, 108, 135003
%
%
%
%
%
\bibitem[Hudson et al. (2001)]{hud01} Hudson, H. S., Kosugi, T., Nitta, N. V., \& Shimojo, M., 2001, \apj, 561, L211
%
\bibitem[Isobe et al. (2002)]{iso02} Isobe, H., Yokoyama, T., Shimojo, M., et al., 2002, \apj, 566, 528
%
\bibitem[Ji\v{r}i\v{c}ka et al. (1993)]{jir93} Ji\v{r}i\v{c}ka, K., Karlick\'{y}, M., Kepka, O., \& Tlamicha, A., 1993, \solphys, 147, 203
%
%
\bibitem[Karlick\'{y} et al. (2002)]{kar02} Karlick\'{y}, M., F\'{a}rn\'{i}k, F. \& M\'{e}sz\'{a}rosov\'{a}, H., 2002, A\&A, 395, 677
%
\bibitem[Karlick\'{y} (2004)]{kar04a} Karlick\'{y}, M. 2004, A\&A, 417, 325
%
\bibitem[Karlick\'{y} et al. (2004)]{kar04b} Karlick\'{y}, M., F\'{a}rn\'{i}k, F., \& Krucker, S., 2004, A\&A, 419, 365
%
%
%
\bibitem[Karlick\'{y} \& B\'{a}rta (2011)]{kar11} Karlick\'{y}, M., \& B\'{a}rta, M., 2011, \apj, 733, 107
%
%
%
\bibitem[Khan et al. (2002)]{kha02} Khan, J. I., Vilmer, N., Saint-Hilaire, P., \& Benz, A. O., 2002, A\&A, 388, 363
%
%
%
\bibitem[Kliem et al. (2000)]{kli00} Kliem, B. Karlick\'{y}, M., \& Benz, A. O., 2000, A\&A, 360, 715
%
\bibitem[Koloma\'{n}ski \& Karlick\'{y} (2007)]{kol07} Koloma\'{n}ski, S. \& Karlick\'{y}, M., 2007, A\&A, 475, 685
%
\bibitem[Krucker et al. (2010)]{kru10} Krucker, S., Hudson, H. S., Glesener, L., et al. 2010, \apj, 714, 1108
%
\bibitem[Kumar \& Cho (2013)]{kum13} Kumar, P. \& Cho, K.-S., 2013, A\&A, 557, A115
%
\bibitem[Kundu et al. (2001)]{kun01} Kundu, M. R., Nindos, A., Vilmer, N., et al., 2001, \apj, 559, 443
%
%
\bibitem[Lin \& Forbes (2000)]{lin00} Lin, J. \& Forbes, T. G., 2000, JGR, 105, 2375	
%
\bibitem[Lin et al. (2002)]{lin02} Lin, R. P., Dennis, B. R., Hurford, G. J., et al., 2002, \solphys, 210, 3
%
\bibitem[Liu et al. (2010)]{liu10} Liu, R., Lee, J., Wang, T., et al., 2010, \apj, 723, L28
%
\bibitem[Liu et al. (2013)]{liu13} Liu, W., Chen, Q., \& Petrosian, V., 2013, \apj, 767, 168
%
%
%
%
%
%
\bibitem[Milligan et al. (2010)]{mil10} Milligan, R. O., McAteer, R. T. J., Dennis, B. R., \& Young, C. A., 2010, \apj, 713, 1292
%
%
\bibitem[Narukage \& Shibata (2006)]{nar06} Narukage, N. \& Shibata, K., 2006, \apj, 637, 1122
%
\bibitem[Neupert (1968)]{neu68} Neupert, W. M., 1968, \apj, 153, L59
%
%
%
%
%
\bibitem[Nishida et al. (2013)]{nisd13} Nishida, K., Nishizuka, N., \& Shibata, K., 2013, \apj, 775, L39
%
\bibitem[Nishizuka et al. (2010)]{nis10} Nishizuka, N., Takasaki, H., Asai, A., \& Shibata, K., 2010, \apj, 711, 1062
%
\bibitem[Nishizuka \& Shibata (2013)]{nis13} Nishizuka, N. \& Shibata, K., 2013, PRL, 110, 051101
%
\bibitem[Ohyama \& Shibata (1997)]{ohy97} Ohyama, M., \& Shibata, K., 1997, PASJ, 49, 249
%
%
\bibitem[Oka et al. (2010)]{oka10} Oka, M., Phan, T.-D., Krucker, S., et al., 2010, \apj, 714, 915
%
\bibitem[Ono et al. (2011)]{ono11} Ono, Y., Hayashi, Y., Ii, T., et al., 2011, Phys. Plasmas 18, 111213
%
\bibitem[Priest (1985)]{pri85} Priest, E. R., 1985, Reports on Progress in Physics (ISSN 0034-4885), 48, 955
%
%
%
%
\bibitem[Reeves \& Golub (2011)]{ree11} Reeves, K. K., \& Golub, L., 2011, \apj, 727, L52
%
\bibitem[Reid et al. (2011)]{rei11} Reid, H. A. S., Vilmer, N., \& Kontar, E. P., 2011, A\&A, 529, A66
%
%
\bibitem[Savage et al. (2010)]{sav10} Savage, S. L., McKenzie, D. E., Reeves, K. K., Forbes, T. G., \& Longcope, D. W., 2010, \apj, 722, 329
%
%
%
\bibitem[Shibata et al. (1995)]{shi95} Shibata, K., Masuda, S., Shimojo, M., et al., 1995, \apj, 451, L83
%
\bibitem[Shibata \& Tanuma (2001)]{shi01} Shibata, K., \& Tanuma, S., 2001, Earth Planets Space, 53, 473
%
%
\bibitem[Shimizu et al. (2009)]{shim09} Shimizu, T., Kondoh, K., Shibata, K., \& Ugai, M., 2009, Phys. Plasmas, 16, 052903
%
%
%
\bibitem[Sui et al. (2003)]{sui03} Sui, L., \& Holman, G. D., 2003, \apj, 596, L251
%
%
\bibitem[Takasao et al. (2012)]{tak12} Takasao, S., Asai, A., Isobe, H., \& Shibata, K., 2012, \apj, 745, L6
%
%
\bibitem[Tan et al. (2007)]{tan07} Tan, B., Yan, Y., Tan, C., \& Liu, Y., 2007, \apj, 671, 964
%
\bibitem[Tan (2008)]{tan08} Tan, B., 2008, \solphys, 253, 117
%
\bibitem[Tan et al. (2008)]{tang08} Tan, C., Yan, Y. H., Liu, Y. Y., et al., 2008, Advances in Space Research, 41, 969
%
\bibitem[Tanaka et al. (2010)]{tan10} Tanaka, K. G., Yumura, T., Fujimoto, M., et al., 2010, Phys. Plasmas, 17, 102902
%
\bibitem[Temmer et al. (2008)]{tem08} Temmer, M., Veronig, A. M., Vr\v{s}nak, B., et al., 2008, \apj, 673, L95
%
%
%
\bibitem[Tsuneta et al. (1997)]{tsu97} Tsuneta, S., Masuda, S., Kosugi, T., \& Sato, J., 1997, \apj, 478, 787
%
%
%
\bibitem[Wang et al. (2012)]{wan12} Wang, R., Tan, B., Tan, C., \& Yan, Y., 2012, \solphys, 278, 411
%
\bibitem[Webb et al.  (2003)]{web03} Webb, D. F., Burkepile, J., Forbes, T. G. \& Riley, P., 2003, JGR, 108, 1440 
%
\bibitem[Wild et al. (1954)]{wil54} Wild, J. P., Roberts, J. A. \& Murray, J. D., 1954, Nature, 173, 532
%
\bibitem[Yokoyama et al. (2001)]{yok01} Yokoyama, T., Akita, K., Morimoto, T., et al., 2001, \apj, 546, L69
%
%
\bibitem[Zhang et al. (2001)]{zha01} Zhang, J., Dere, K. P., Howard, R. A., et al., 2001, \apj, 559, 452
%
%
%
\end{thebibliography}
\end{document}